\documentclass[aps,prb,twocolumn,reprint,amsmath,amssymb,superscriptaddress, showpacs]{revtex4}
\usepackage{graphicx}% Include figure files
\usepackage{dcolumn}% Align table columns on decimal point
\usepackage{bm}% bold math
\usepackage{amsmath}
\usepackage{amssymb,amsthm}
\usepackage{color}
\usepackage[sans]{dsfont}
\usepackage{subfig}
\usepackage{epstopdf}
\usepackage{mathbbol,bbm}

\let\originalleft\left
\let\originalright\right
\renewcommand{\left}{\mathopen{}\mathclose\bgroup\originalleft}
\renewcommand{\right}{\aftergroup\egroup\originalright}

\newcommand{\<}{\langle}
\renewcommand{\>}{\rangle}

\def\r{\hat{\rho}}

\def\ham{\hat{\mathcal{H}}}
\def\partition{\mathcal{Z}}

\def\pauz{\hat{\sigma}^z}
\def\paux{\hat{\sigma}^x}

\def\fkik{F_{\small\mbox{Kik}}}
\newcommand{\tr}[2]{\text{Tr}_{#1}[#2]}

\begin{document}
\title{The Quantum Cluster Variational Method and the Phase Diagram of the quantum ferromagnetic $J_1$-$J_2$ model}
\author{E. Dom\'{\i}nguez} 
\email{eduardo@fisica.uh.cu}
\affiliation{Group of Complex Systems and Statistical Physics, Department of Theoretical Physics, University of Havana, Cuba}

\author{C.E.Lopetegui} 
\email{eduardo@fisica.uh.cu}
\affiliation{Group of Complex Systems and Statistical Physics, Department of Theoretical Physics, University of Havana, Cuba}

\author{Roberto Mulet} 
\email{mulet@fisica.uh.cu}
\affiliation{Group of Complex Systems and Statistical Physics, Department of Theoretical Physics, University of Havana, Cuba}

\date{\today}
%\pacs{05.60.Gg, 03.65.Xp, 72.10.-d, 87.15.hj}

%Quantum transport (general physics)
%Tunneling (general physics)
%Theory of electronic transport (cond mat)
%Transport dynamics (biomolecules)

\begin{abstract}
  We exploit the Quantum Cluster Variational Method (QCVM) to study the $J_1$-$J_2$ model for quantum Ising spins. We first describe the QCVM and discuss how it is related to other Mean Field approximations. The phase diagram of the model is studied at the level of the Kikuchi approximation in square lattices as a function of the ratio between $g=J_2/J_1$, the temperature and the longitudinal and transverse external fields. Our results show that quantum fluctuations may change the order of the transition and induce a gap between the ferromagnetic and the stripe phases. Moreover, when both longitudinal and transverse fields are present thermal fluctuations and quantum effects contribute to the appearance of a nematic phase.
\end{abstract}

\maketitle

\section{Introduction}

In most cases of practical interest the solution of problems involving many interacting quantum particles rests on some kind of approximation \cite{SS,CMP1,CMP2}. It is not surprising then that an important fraction of the theoretical work in Condensed Matter Physics is concerned with the development of such approximations and the test of the predictions derived from them. Many of these techniques are similar in spirit, they improve over the simplest Mean Field approximation to the problem including correlations, and/or interactions between clusters of variables.  Among the most celebrate approximation, although certainly not the only ones, we could mention, the Effective Field Theory (EFT) \cite{Bobak18}, the Cluster Mean Field (CMF) \cite{Jin13,Ren14} and the Correlated Cluster Mean Field (CCMF) \cite{Yamamoto,Zimmer,Singhania18}. The Quantum Cluster Variational method (QCVM)\cite{ourQCVM}, inserts into this family in a novel way: it allows the presence of disorder in the model, establishing a clear distinction between average case scenarios and computations on single instances, and connecting with message passing algorithms developed in Computer Science and Information Theory\cite{MeMo}.

QCVM has its roots in previous results from Morita and Tanaka \cite{Morita94a,Morita94b,Tanaka94} that starting from a pure variational approximation to the free energy of finite dimensional systems derived a set of close equations for the order parameters of a quantum problem defined by a Hamiltonian. Within this formulation, the resulting set of equations is usually solved assuming the existence of specific symmetries for the order parameter of the model. QCVM goes a step further, extending an approach previously developed for classical disordered models \cite{tommaso_CVM,GBP_GF,RCVM} to connect these variational approximations with message passing equations in finite dimensional systems. The solutions of these equations can be computed either self-consistently in specific instances of the problem or on average over the disorder\cite{ourQCVM,RCVM}. QCVM also generalizes, to finite dimensional lattices, the work done by \cite{Evans2008,Leifer2008,Poulin2008,Bilgin2009,Poulin2011,Farzad2014,Biazzo2013,Biazzo2014}, that used message passing algorithms to solve quantum disordered problems in systems with tree-like topologies.

In this work we exploit the QCVM to study the phase diagram of the quantum $J_1$-$J_2$ model in the presence of external fields. This model belongs to
an extensively studied family of similar systems where competing interactions induce exotic phases at low temperatures. This includes systems with short range ferromagnetic and long range dipolar interactions \cite{magnfilms,Rogelio,Fey16,Fey19,AMC15,AMC17,AMC20,cannas2006}, long range Coulomb interactions \cite{Nandkishore17} and two-dimensional dipolar Fermi gases\cite{Oganesyan07,Bruun08,Parish12}.

The  $J_1$-$J_2$ model has captured special attention both because of its apparent simplicity, and because of its resemblance with real materials. The classical version has been studied extensively\cite{Moran1993,Moran1994,Stariolo,Ours,DOSANJOS2008,Kalz2009,Kalz2011,Jin2012} but the quantum version has turned harder to crack. Mainly, because quantum fluctuations add complexity to the inherent frustration of the model inducing a zoology of exotic phases: Stripes like phases, columnar anti-ferromagnetic (CAF), N\'eel Anti-ferromagnetic (NAF) phases, Spin-liquid phases and Spin Nematic Phases have been theoretically predicted \cite{Kohama2019, Shannon2004,Mezzacapo2012,Jiang2012,Stariolo,Dagotto1989,Shindou2013,Cysne_2015}. Some of them, like stripes and N\'eel Antiferromagnets, have been experimentally observed \cite{Carretta2002,Melzi2001,Nath2008} while others, like the Spin Nematic phase and the Spin-liquid phase lack of conclusive experimental observations but keep the interest of experimentalists \cite{Shannon2004,Orlova2017}.

Our intention in this work is twofold. On one hand to test the Quantum Cluster Variational Method in a quantum model with competing interactions. On the other, to unveil the effect of quantum fluctuations on the phase diagram of the ferromagnetic $J_1$-$J_2$ model with competitive interactions. We study this model extensively as a function of the ratio $g=J_2/J_1$, the temperature $T$ and external fields. We show that the approximation captures a rich  phenomenology of phases and phase transitions.

The work is organized as follows. In the next section we present the model and summarize some of the known results of the literature. We continue introducing the Quantum Cluster Variational Method  and how it applies to the $J_1$-$J_2$ model. In this section we also discuss how QCVM is connected with other Mean Field Methods in the literature. Then, we show and discuss the results of our computations for the $J_1$-$J_2$ model. In the last section we present the conclusions of our work.

\section{The Model}

The quantum $J_1$-$J_2$ model is one of the most studied models of magnetic materials displaying frustration and is defined by the Hamiltonian
\begin{equation}
 \ham= -\sum_{\langle ij\rangle}J_{1}\hat{\sigma}_{i}^{x}\hat{\sigma}_{j}^{x}-\sum_{\langle\langle ij\rangle\rangle}J_{2}\hat{\sigma}_{i}^{x}\hat{\sigma}_{j}^{x}-\sum_{\left(i\right)} \vec{h_{i}} \cdot \vec{\sigma}_{i}, 
\end{equation}
where $\langle ij\rangle$ stand for the Nearest Neighbors in the square lattice and $\langle\langle ij\rangle\rangle$ for the Next Nearest Neighbors and the external field $\vec{h_i}$ may point to an arbitrary direction. We focus here our attention in the model with nearest-neighbour interaction $J_1$ greater than zero (ferromagnetic), and $J_2\leq 0$. The later induces the frustration into the model. Since the preferred direction of the system is labeled \textit{x}, quantum effects appear when the transverse field $h_z$ is different from zero.

The classical model ($h_z=0$) has been largely studied\cite{Moran1993,Moran1994,Stariolo,Ours,DOSANJOS2008,Kalz2009,Kalz2011,Jin2012}. However, also there some questions remain open. For example, while for values of $g=\frac{|J_2|}{J_1}$ slightly greater than $0.5$ the transition is discontinuous, Monte Carlo studies\cite{Kalz2009,Kalz2011} suggest a continuous transition for $g \sim 1$. In a paper by Jin et al.\cite{Jin2012}, the existence of a pseudo first order transition in the range $g_c\leq g\lesssim 1$ is reported, with $\,g_c=0.67$. They found that the critical exponents vary continuously between those of the 4-state Potts model at $g=g_c$ to those of the Ising model for $g\rightarrow \infty$. 

Recently, researchers paid attention also to the effect of external magnetic fields and demonstrated the presence of a nematic phase\cite{Stariolo,Ours}, with slight differences, at low temperatures, between the two studies. This nematic phase is characterized by the presence of orientational order and the lack of positional order and has been reported experimentally \cite{Orlova2017}. 

In the present work we report, exploiting QCVM, our own perception of this problem, and show that there is no clear separation between pseudo and actual first order behavior. Moreover, we studied the combined effect of transverse and longitudinal field to understand the role played by quantum or thermal fluctuations in the order of these transitions.

\begin{figure}[!htb]
\centering
\includegraphics[width=.4\textwidth]{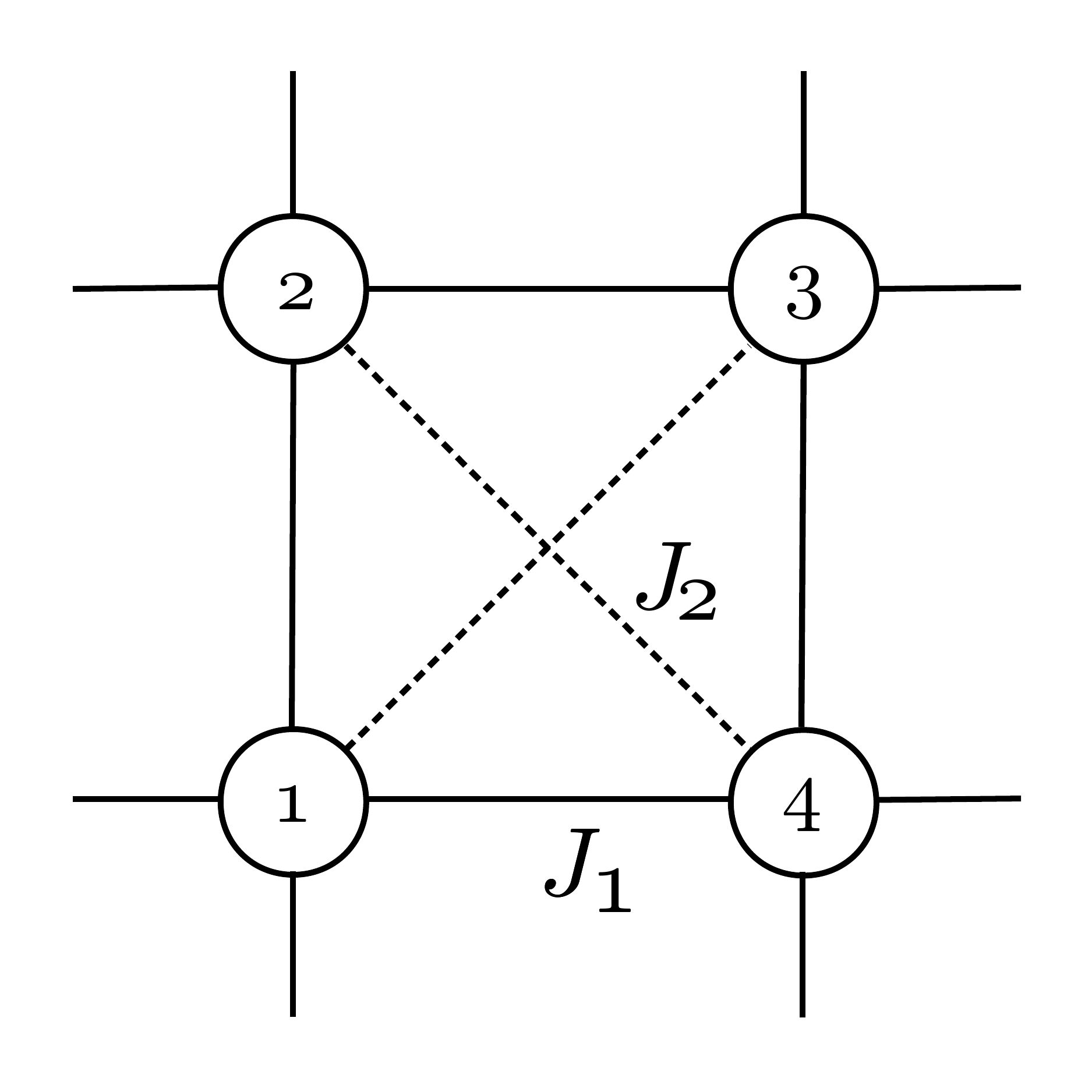}%phase_diagram_Ferro_J1J2.png}
\caption{Convention used in the definition of the order parameters. }
\label{numeration}
\end{figure}

We explore the behaviour of three order parameters\cite{Stariolo,Ours}. First, we define two positional order parameters which measures the breaking in the translational symmetry of the lattice in two possible directions:
\begin{equation}
 M_x=\frac{|m^{x}_1-m^{x}_4|}{2}
\end{equation}
\begin{equation}
 M_y=\frac{|m^{x}_1-m^{x}_2|}{2}
\end{equation}
Where $m^{x}_i$ refers to the mean magnetization along the preferred direction of the system, of the site \textit{i}, inside a given plaquette, following the convention showed in Fig.\ref{numeration}.

While translational order parameters are enough to determine the presence of stripes, they are not suited to detect the breaking of the rotational symmetry of the lattice. They take the same values in a paramagnetic and a nematic phase, yet the nematic phase shows a structure of the correlations similar to that of the stripes. This is natural if we think about this phase as an intermediate one in between the completely symmetric paramagnetic phase and the ordered stripes phase. Therefore, to characterize the nematic behavior we introduced an orientational order parameter defined as
\begin{equation}
 Q=\frac{|l_{12}+l_{34}-l_{14}-l_{23}|}{4},
\end{equation}
where $l_{ij}$ represents the correlation between sites $i$ and $j$ in the plaquette. All the order parameters are defined in a way that $0<\{Q,M_x,M_y\}<1$.

\section{Mean Field Approximations and Quantum Cluster Variational Method}

In order to compute the order parameters relevant for the $J_1$-$J_2$ model
it is fundamental to obtain a good approximation of at least local magnetizations and 
correlations of neighboring spins. These quantities can be found from a knowledge
of the local density matrices describing the joint statistics of these spins.

With a cluster expansion the free energy is approximated as a sum of contributions of selected regions or groups of variables. Each term of the sum depends on the local density matrix of the corresponding variables (spins). These regions are typically chosen to balance the accuracy and  the computational cost of the calculations. The resulting 
approximated expression is to be understood as variational in terms of the local densities. The quality of the final results depends directly on the choice of regions, their size in relation to the correlation length, and how much mean-field is put the treatment of effective interactions and correlation between regions.

Among the many cluster approximations appearing in the literature, the approach known as Cluster Variational Method or CVM is remarkable because of the systematic treatment of different levels of complexity \cite{YFW05,kikuchi}. 
In a natural way one can choose free energy expansions ranging from naive mean field, a Bethe approximation or more involved structures. A quantum extension of such ideas, QCVM, was developed in \cite{Morita94a,Morita94b,Tanaka94}  and more recently expanded in \cite{ourQCVM}. 
In particular we are interested in a construction called Kikuchi approximation. The Kikuchi free energy $\fkik$ consists of a sum of local free energies of all the square unit cells ($p$), the interacting pairs ($l$) formed by the overlap of neighbor cells\footnote{The diagonal interactions do not belong to any overlap region and therefore, according to the CVM, they do not appear explicitly as independent terms in the approximated free energy.} and all the contributions from single spins ($s$):

\begin{equation}
 \fkik = \sum_{p} c_p F_p + \sum_{l} c_l F_l+ \sum_{s}c_s F_s
\end{equation}

The local free energy terms are defined the usual way
\begin{eqnarray}
 F_r = \tr{}{\ham_r \r_r} + T \tr{}{\r_r \ln \r_r}
\end{eqnarray}
where $r$ may label a plaquette, link or spin region. The numbers $c_p,c_l,c_s$
are weight factors that compensate the effect of overlap between regions. Due
to overlaps, the contribution of some variables might be under or over represented if the weight factors were absent. For the case of interest here,
$c_p = 1, c_l = -1, c_s = 1$. 

The minimization of the functional $\fkik$ should be done under certain constraints. In addition to the usual normalization conditions, $\tr{}{\r_r} = 1$, the consistency between different regions must be enforced explicitly. 
For all plaquettes $p = \<ijkm\>$ and any of its forming links $l = \<ij\>$ one could demand that the partial trace over spins $k$ and $m$ equals the link density $\r_{l}$:
\begin{equation}
 \tr{p\setminus l}{\r_p} = \r_{l}.
 \label{eqn:hard_plaquette_link_constraint}
\end{equation}
Moreover, every link distribution $\r_{l}$ could be forced to marginalize onto the corresponding spin densities $\r_i$ and $\r_j$:
\begin{equation}
 \tr{j}{\r_l}= \r_{i} \quad ; \quad\tr{i}{\r_l}= \r_{j}.
\end{equation}
However, numerical results\cite{ourQCVM} have shown that the above constraint scheme
is too restrictive. It turns out that the minimization gives much more accurate
results if consistency is enforced only in the direction of the spin-spin interactions. In this model, where interactions exist only in the OX direction, it amounts to relaxing \eqref{eqn:hard_plaquette_link_constraint} to the set of equations:

\begin{equation}
 \left.
 \begin{array}{c}
 \tr{}{\r_p \paux_i \paux_j} = \tr{}{\r_{l} \paux_i \paux_j} \\ \\
 \tr{}{\r_p \paux_i} = \tr{}{\r_{l} \paux_i} \\ \\
 \tr{}{\r_p \paux_j} = \tr{}{\r_{l} \paux_j}
 \end{array}
 \right\rbrace 
 \quad \forall \; p, \forall \; l = \<ij\> \in p
 \label{eqn:soft_plaquette_link_constraint}
\end{equation}

\begin{equation}
 \left.
 \begin{array}{c}
 \tr{}{\r_l \paux_i} = \tr{}{\r_i \paux_i} \\ \\
 \tr{}{\r_l \paux_j} = \tr{}{\r_j \paux_j}
 \end{array}
 \right\rbrace 
 \quad \forall \; l = \<ij\>
 \label{eqn:soft_link_spin_constraint}
\end{equation}

We are now in position to write a Lagrange function that 
includes the conditions \eqref{eqn:soft_plaquette_link_constraint} and \eqref{eqn:soft_link_spin_constraint} by means of suitable Lagrange multipliers $\hat \lambda$, $\hat \gamma$:

\begin{eqnarray}
 L[\{\r_p\},\{\r_l\},\{\r_s\}] &=& \fkik \nonumber \\
 &+& \sum_p \sum_{l\in p} \tr{l}{\hat\lambda_{p \rightarrow l} \left(\r_l - \tr{p\setminus l}{\r_p}\right)} \nonumber\\
 &+& \sum_l \sum_{s\in l} \tr{s}{\hat\gamma_{l \rightarrow s}\left(\r_s - \tr{l\setminus s}{\r_l}\right)}\nonumber\\
 &+& \text{normalization conditions}
\end{eqnarray}

To conform to \eqref{eqn:soft_plaquette_link_constraint} and \eqref{eqn:soft_link_spin_constraint}, it suffices to introduce the parametrization\footnote{Notice that $\lambda$ and $\gamma $ are actually operators on the Hilbert space
of the pairs $l$ and the single spins $s$, respectively.}:

\begin{eqnarray}
 \label{eqn:parametrization_lambda}
 \hat\lambda_{p \rightarrow l} &=& C_{p\rightarrow l} \paux_i \paux_j + c_{p\rightarrow i} \paux_i + c_{p\rightarrow j} \paux_j \\
 \label{eqn:parametrization_gamma}
 \hat\gamma_{l\rightarrow i} &=& d_{l\rightarrow i} \paux_i
\end{eqnarray}
It is enough to make $C,c,d$ all real to guarantee that $\hat \lambda$ and $\hat\gamma$ are Hermitian, and consequently, that $L$ is real. Moreover, this parametrization ensures that the operators $\r$ that satisfy the stationary condition are all positive, which is the fundamental property of density matrices (in addition to an unit trace, of course). 

We will say that the value of the Lagrange function $L[\r_r]$ is stationary with respect to a change in the density $\r_r$ if the linear part of the expansion
of $L[\r_r + \epsilon \delta \r_r]$, in powers of $\epsilon$, is zero. The operator $\delta \r_r$ is arbitrary but should be Hermitian, traceless and small enough to keep $\r_r + \epsilon \delta \r_r$  as a valid density operator,  The Lagrange function will be stationary with respect to the set of all densities
$\{\r_p\},\{\r_l\},\{\r_s\}$, by definition, if it is stationary in each one of them. The stationarity condition can be summarized in the following formula:
\begin{equation}
 \left.\dfrac{\partial L[\r_r + \epsilon \delta \r_r]}{\partial \epsilon}\right|_{\epsilon = 0} = 0 \quad\quad \forall \; \text{region} \;r
 \label{eqn:stationary_condition}
\end{equation}

As an example, let us work out the form of the density matrix of an arbitrary spin $s_0$ at the stationary point. The part of $L$ that depends on a given $\r_{s_0}$ is
\begin{eqnarray}
& \tr{}{\r_{s_0}\ham_{s_0}}& + T\; \tr{}{\r_{s_0} \ln \r_{s_0}} + \alpha_{s_0}(\tr{}{\r_{s_0}} - 1)\nonumber \\ & & +\sum_{l \ni s_0}  \tr{}{\hat\gamma_{l \rightarrow s_0}\left(\r_{s_0} - \tr{l\setminus s_0}{\r_l}\right)} 
 \label{eqn:linear_lagrange_1}
\end{eqnarray}

In \eqref{eqn:linear_lagrange_1} the only term that is somewhat involved to expand to first order in $\epsilon$, when evaluating $L[\r_{s_0} + \epsilon \delta \r_{s_0}]$, is the one with the logarithm. The trick is to use perturbation theory to write the eigenvalues $r_i(\epsilon)$ of $\r_{s_0} + \epsilon \delta \r_{s_0}$ using the eigenvalues and eigenvectors $\{r_i,|r_i\rangle\}$ of $\r_{s_0}$:
\begin{eqnarray}
 r_i(\epsilon) &=& r_i + \epsilon \;(\delta\r_{s_0})_{ii} + o(\epsilon)\\
 (\delta\r_{s_0})_{ii} &=& \langle r_i|\delta\r_{s_0}|r_i\rangle
\end{eqnarray}
The trace with the logarithm is therefore written as
\begin{eqnarray}
 & &\tr{}{(\r_{s_0} + \epsilon \delta \r_{s_0}) \ln (\r_{s_0} + \epsilon \delta \r_{s_0})} = \sum_i r_i(\epsilon) \ln r_i(\epsilon) \nonumber \\
 &=& \sum_i r_i \ln r_i +  \epsilon \sum_i (\delta\r_{s_0})_{ii} (\ln r_i + 1) + o(\epsilon)\nonumber \\
 &=& \tr{}{\r_{s_0}\ln \r_{s_0}} + \epsilon \,\tr{}{\delta\r_{s_0} \ln \r_{s_0}} + o(\epsilon) 
 \label{eqn:last_equation}
\end{eqnarray}

In \eqref{eqn:last_equation} we used that $\delta\r_{s_0}$ has a null trace. From the results above it is not hard to deduce the linear expansion of 
$L[\r_{s_0} + \epsilon \delta \r_{s_0}]$:

\begin{eqnarray}
  & &L[\r_{s_0} + \epsilon \delta \r_{s_0}]  = L[\r_{s_0}] + +\epsilon 
 \tr{}{\delta \r_{s_0}\ham_{s_0}} \nonumber \\
  & &  +\epsilon 
 \tr{}{\delta \r_{s_0}\left( T \ln \r_{s_0} 
    +  \sum_{l\ni s_0}\hat\gamma_{l \rightarrow s_0} \right)} + o(\epsilon)
 \label{eqn:linear_expansion_lagrange}
\end{eqnarray}

The stationarity condition applied to \eqref{eqn:linear_expansion_lagrange} implies the following relation
\begin{equation}
 \tr{}{\delta \r_{s_0}\left(\ham_{s_0} + T \ln \r_{s_0} + \sum_{l\ni s_0}\hat\gamma_{l \rightarrow s_0} \right)} = 0
 \label{eqn:stationary_condition_2}
\end{equation}

Now we use that $\delta \r_{s_0}$ is an arbitrary traceless Hermitian operator
to state that the part in parenthesis in \eqref{eqn:stationary_condition_2} is
at most a constant times the identity, and therefore, $\r_{s_0}$ must have the form:
\begin{equation}
 \r_{s_0} = \dfrac{1}{\partition_{s_0}} \exp -\beta[\ham_{s_0} +  \sum_{l\ni s_0}\hat\gamma_{l \rightarrow s_0}]
 \label{eqn:belief_spin_1}
\end{equation}
with $\beta = 1/T$. The condition $\tr{}{\r_{s_0}} = 1$ determines the value of $\partition_{s_0}$. The form
of \eqref{eqn:belief_spin_1} resembles a Boltzmann distribution where an effective interaction given by the Lagrange multipliers $\hat\gamma_{l \rightarrow s_0}$ is added to the original local Hamiltonian. 

The same procedure that leads to \eqref{eqn:belief_spin_1} can be used to
find expressions for the rest of the density operators, that is, for pair and plaquette densities. The general structure will be the same: a Boltzmann distribution where the local Hamiltonian is modified by a number of effective interactions. These appear due to the consistency relations between regions and via the corresponding Lagrange multipliers. We skip these derivations and only write the final expressions for the Kikuchi approximation:

\begin{equation}
 \begin{array}{rl}
  \r_{s} &=\dfrac{1}{\partition_{s}} \exp -\beta\;\;[\ham_{s} + \displaystyle\sum_{l\ni s}\hat u_{l \rightarrow s}]\\
  \r_{l} &=\dfrac{1}{\partition_{l}} \exp -\beta\;\;{\large [}
  \ham_{l} + \displaystyle\sum_{p \ni l} \hat U_{p \rightarrow l} 
  + \displaystyle\sum_{\substack{l'\ni i\\ l'\neq l}}\hat u_{l' \rightarrow i} 
  +  \displaystyle\sum_{\substack{l'\ni j\\ l'\neq l}}\hat u_{l' \rightarrow j}]\\
 % \quad\quad \forall \;l = \<ij\> \\
  \r_{p} &=\dfrac{1}{\partition_{p}} \exp -\beta [
  \ham_{p} + \displaystyle\sum_{l\in p}\sum_{\substack{p' \ni l\\p'\neq p}} \hat U_{p' \rightarrow l}
  +  \displaystyle\sum_{i \in p}\sum_{\substack{l'\ni i\\ l'\notin p}}\hat u_{l' \rightarrow i}]
 \end{array}
 \label{eqn:belief_set_equations}
\end{equation}

In \eqref{eqn:belief_set_equations} we used  a linear transformation
to write $\hat \lambda_{p \rightarrow l}$ and $\hat \gamma_{l \rightarrow i}$
as a function of new parameters $\hat U_{p \rightarrow l}$ and $\hat u_{p \rightarrow i}$, respectively. These changed variables will of course
inherit the structure of the originals, see \eqref{eqn:parametrization_lambda}
and \eqref{eqn:parametrization_gamma}:
\begin{equation}
\begin{array}{rcl}
 \hat U_{p \rightarrow l} &=& - U_{p \rightarrow l} \paux_i \paux_j - 
 u_{p \rightarrow i} \paux_i - u_{p \rightarrow j} \paux_j\\
 \hat u_{l \rightarrow s} &=& - u_{l \rightarrow s} \paux_s
 \end{array}
 \label{eqn:parametrization_u}
\end{equation}

The form of \eqref{eqn:parametrization_lambda}, \eqref{eqn:parametrization_gamma} and \eqref{eqn:parametrization_u} suggests that the Lagrange multiplier can be interpreted as effective local fields 
and a contribution to the  interaction constants between spins when their explicit form is put into the equations \eqref{eqn:belief_set_equations}. The plaquette-to-pair 
$\hat U_{p \rightarrow l}$ is composed of a terms that adds to the $J_1 \paux_i \paux_j$  part of the original local Hamiltonian, and two effective fields
$u_{p \rightarrow i},u_{p \rightarrow j}$ acting on the spins of the pair $l = \langle ij\rangle$. On the other hand, the pair-to-spin $\hat u_{l \rightarrow s}$
is formed by an effective field on $\paux_s$, for every spin $s$ that belongs
to $l$. The role of these extra fields is to guarantee
the consistency between the distributions, according to the restrictions
imposed to the variational function. 

The nature of the constraints directly implies that no effective fields (or modification to interaction constants) act in the transverse (OZ) direction.
The reason being that no consistency is demanded on this axis. As a consequence, moments and correlations in the OZ direction may differ if computed from different local densities. Since all the observables we
study here are in OX, where consistency holds, this discrepancy is not
a serious complication.

\begin{figure}[!htb]
\centering
\includegraphics[width=.4\textwidth]{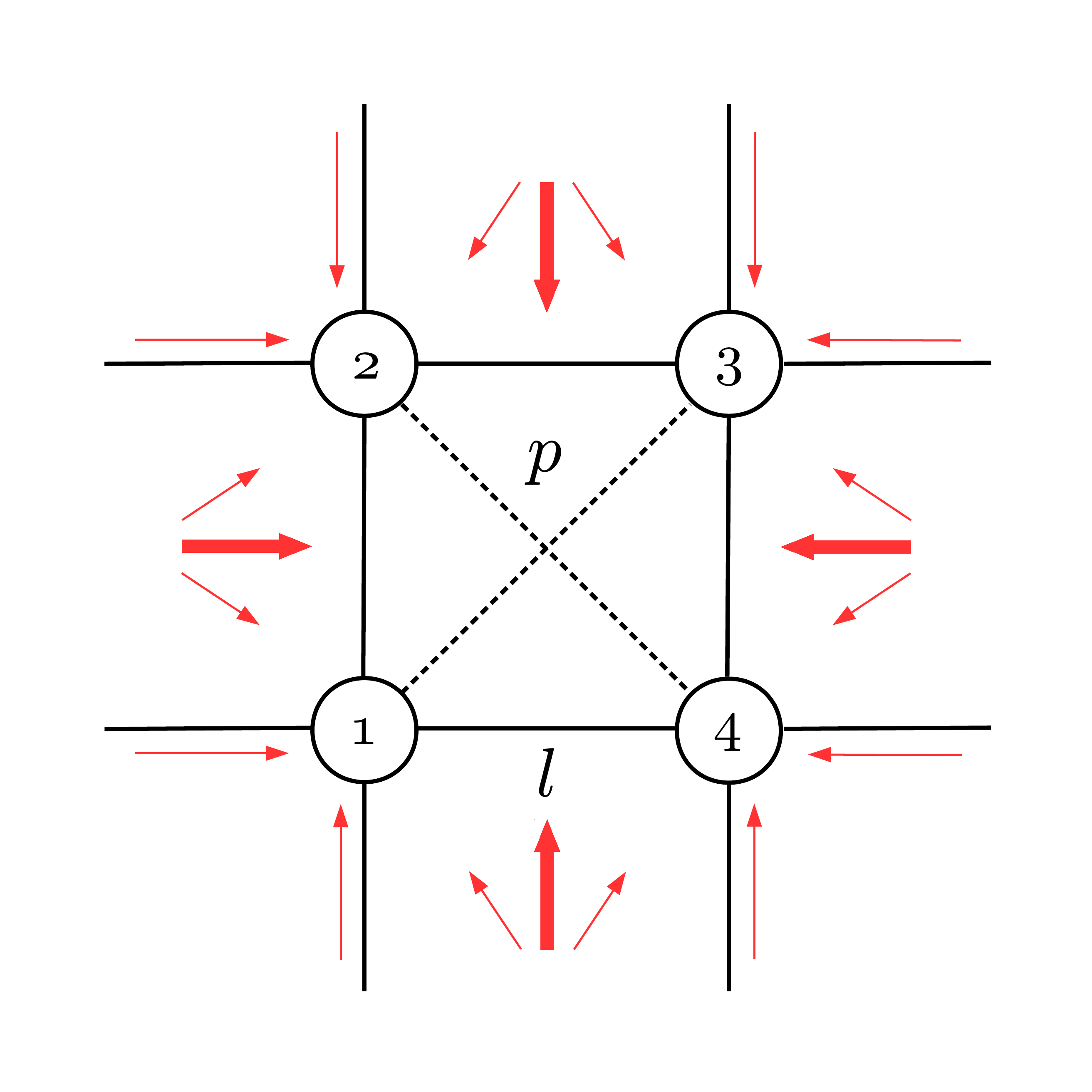}%phase_diagram_Ferro_J1J2.png}
\caption{Effective fields involved in the computation of $\r_p$ for a
plaquette $p$, according to formula \eqref{eqn:belief_set_equations}. Diagonal interactions for plaquettes other than $p$ are not
shown for simplicity.}
\label{fig:kikuchi_fields}
\end{figure}

\begin{figure}[!htb]
\centering
\includegraphics[width=.4\textwidth]{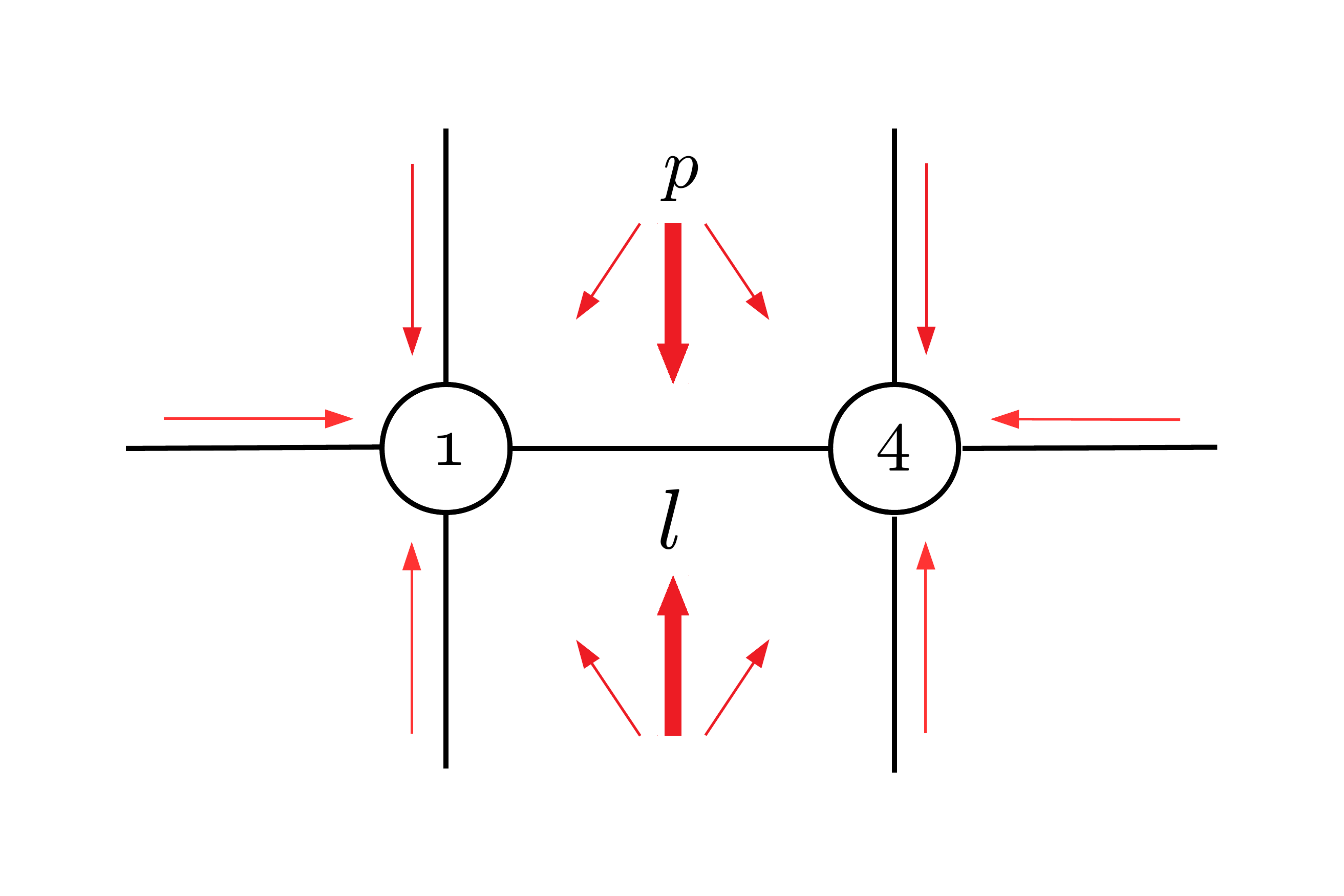}%phase_diagram_Ferro_J1J2.png}
\caption{Effective fields involved in the computation of $\r_l$ for a
pair $l$, according to formula \eqref{eqn:belief_set_equations}. Diagonal
interactions are omitted for simplicity.}
\label{fig:bethe_fields}
\end{figure}

\begin{figure}[!htb]
\centering
\includegraphics[width=.4\textwidth]{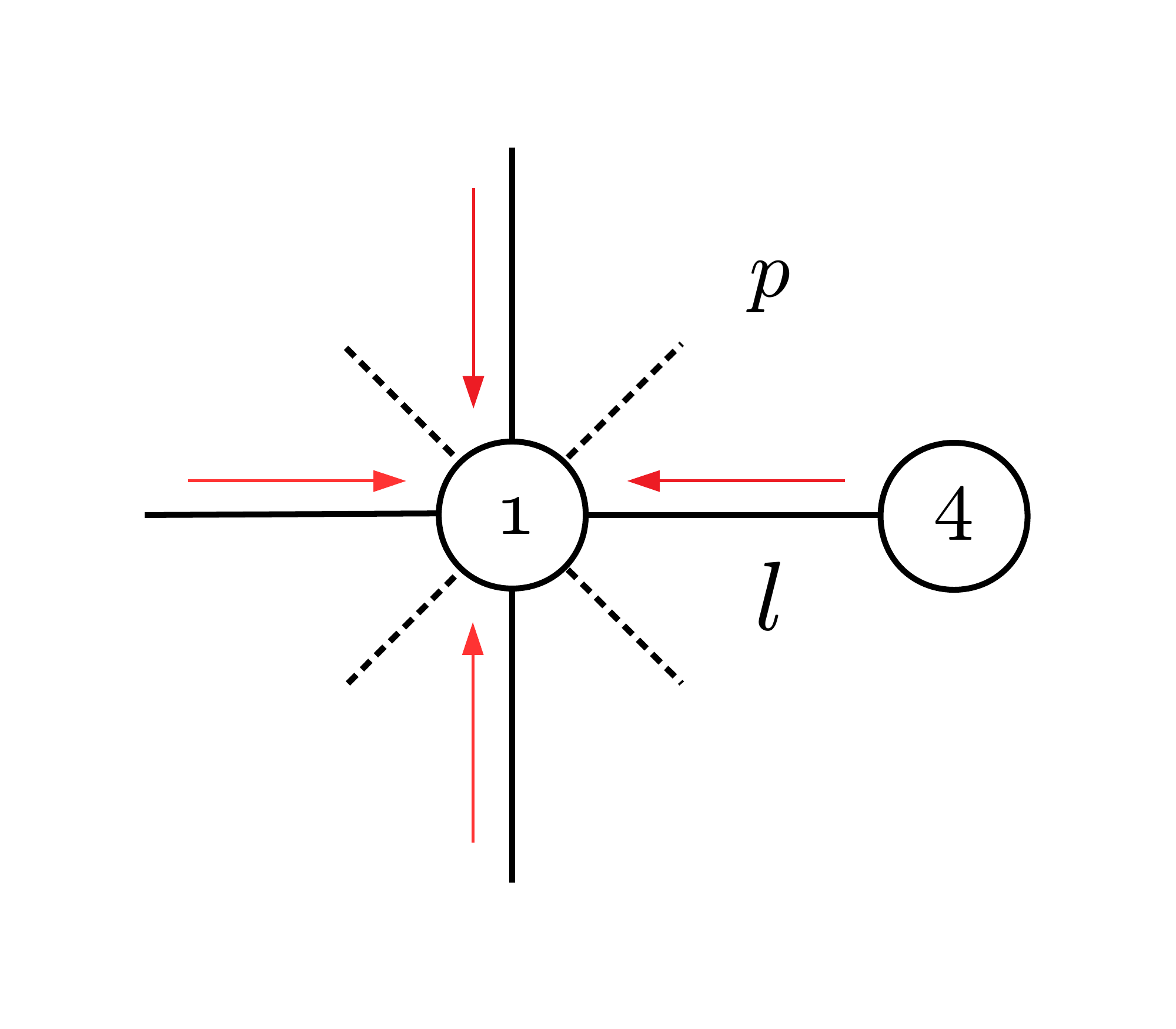}%phase_diagram_Ferro_J1J2.png}
\caption{Effective fields involved in the computation of $\r_s$ for a
single spin $s$ ($s = 1$ in this case), according to formula \eqref{eqn:belief_set_equations}. Diagonal interactions are represented only
for reference, but no effective field appears on these edges.}
\label{fig:spin_fields}
\end{figure}

A graphical representation of the effective fields entering the formula \eqref{eqn:belief_set_equations}
for the plaquette density operator $\r_p$ is shown in Figure (\ref{fig:kikuchi_fields}). Effective fields are drawn as arrows, placed
according to the regions they are related to. For example, the arrows standing parallel to the edges
of the lattice (NN interactions) correspond to $\hat u_{l' \rightarrow i}$
terms. The arrows (triplet) pointing from the center of a square to a border pair picture the $\hat U_{p' \rightarrow l}$ operators. In the same way,
Figures (\ref{fig:bethe_fields}) and (\ref{fig:spin_fields}) provide the 
representation for distributions $\r_l$ and $\r_s$.

The final step to determine the local distributions is to find the value of the effective parameters. Their value is fixed by the solution of the set of equations imposed by the consistency relations \eqref{eqn:soft_plaquette_link_constraint} and \eqref{eqn:soft_link_spin_constraint}. The numerical algorithm used to solve these coupled equations is very similar to the classical generalization of the well known Belief Propagation and is described in detail in \cite{ourQCVM}. 

The basic scheme is the following. All effective fields are initialized
in an arbitrary way (randomly, uniform, following a certain pattern, etc).
Then, a series of update steps are performed until convergence is achieved\footnote{In certain cases the algorithm fails to converge. This will be
discussed afterwards together with the numerical results.}.
We understand convergence as the situation where the values of the effective
fields are such that all the consistency relations are satisfied simultaneously.

Let us describe a typical iteration for the field $\hat U_{p \rightarrow (14)}$ [upper triplet in Figure (\ref{fig:bethe_fields})].
First, compute $\r_p$ using the current value of the fields depicted
in Figure (\ref{fig:kikuchi_fields}). Then, from $\r_p$ determine the correlation $c_{14} = \langle \paux_1 \paux_4\rangle$ and the magnetizations $m_1 = \langle \paux_1\rangle$, $m_4 = \langle \paux_4\rangle$. The crucial part is to use the consistency condition between $\r_p$ and $\r_{14}$. 
We should find a value for the three components of 
$\hat U_{p \rightarrow (14)}$ (this parameter enters the formula for $\r_{14}$ and \textit{not} $\r_p$) such that the  observables computed from the pair distribution $\r_{14}$ match the plaquette predictions:
\begin{align}
 \langle \paux_1 \paux_4\rangle_{\r_{14}} &=  c_{14}\\
 \langle \paux_1\rangle_{\r_{14}} &= m_1 \\
 \langle \paux_1\rangle_{\r_{14}} &= m_4
\end{align}
The three equations above suffice to find the three parameters $\hat U_{p \rightarrow (14)}\sim(U_{p \rightarrow (14)}, u_{p \rightarrow 1},u_{p \rightarrow 4} )$. The newly found value of $\hat U_{p \rightarrow (14)}$ is then updated in the lattice\footnote{Actually, what is usually stored is a weighted
sum of the new and the old values. This procedure introduces a kind of \textit{damping} that might help convergence.}.

Updating a pair-to-spin field, for example $\hat u_{(14) \rightarrow 1}$
follows a similar recipe. First compute $\r_{l=(14)}$ using the fields
of Figure (\ref{fig:bethe_fields}). Then compute the spin 1 magnetization
$\langle \paux_1\rangle$ using $\r_{l}$. Finally, find a value $u_{(14) \rightarrow 1}$ such that the same magnetization computed with $\r_1$ matches the value computed with the pair distribution.

The algorithm described here assumes that we are using a given realization
of the model, that is, that we are computing all the effective fields
for every region defined in a $N = L^2$ square latice. The procedure is general and can handle situations where the external fields applied may not be homogeneous or the interactions change across the lattice. However, since the model in question is translationally invariant, it is reasonable to expect a simplification of the calculations; one would hope this structure is reflected  in a translational symmetry of the effective fields. This is indeed the case. If the same values repeat all over the lattice, it is not necessary to solve $O(N)$ equations but only a reduced
set. The fields outside a certain region of the lattice can be taken as
the same computed inside that region. In this case the numerical procedure looks like 
a self-consistent iteration. The idea, though, is not that every site in the lattice is equivalent to the rest, that is a too restrictive assumption that would allow only homogeneous states. The correct procedure is to consider the smallest possible structures that, by repetition, can create
the patterns observed experimentally. In practice, to generate a pattern
of stripes, or an anti-ferromagnetic checkerboard design, it is enough to
consider a square plaquette as the elementary region. For the model studied here we followed both approaches, a specific realization of the lattice solving $N$ equations, and the more practical reduced version obtained equivalent results. For generality we will present only simulations results for the former method. On one hand it makes clearer the strength of QCVM for more general situations, on the other, makes more transparent when the algorithm does not converge.

To close this section we would like to make a few comments about our choice 
of the QCVM as the inference method used to compute the local observables of the $J_1$-$J_2$ model. Mean field cluster approximations frequently found in the literature \cite{Kellerman2019,Balcerzak2018,Bobak18} are often based on some kind of variational argument on single site spin magnetizations. In many cases the effect is equivalent to factoring out the correlations as products of single site moments. A more consistent treatment is expected to take into account more structure
into the relevant correlations among the system variables. That is precisely
the formal advantage of QCVM: by means of the effective interactions the correlations have extra parameters that can be used for tuning.
This can lead to more precision in the prediction of system properties. A quick 
test, for example, shows that the Cluster Mean Field of \cite{Kellerman2019} applied
to a classical Ising model on a square lattice returns predictions for the para-ferro transition temperature, $T_{CMF}\approx 3.5$ that is closer to the naive MF prediction $T_{MF} = 4.0$ than to
the real value $T_c\approx 2.23$. Bethe-Pierls approximation  or Kikuchi give much more accurate estimates (2.89 and 2.41 respectively).

\begin{figure}[!htb]
\centering
\includegraphics[width=.4\textwidth]{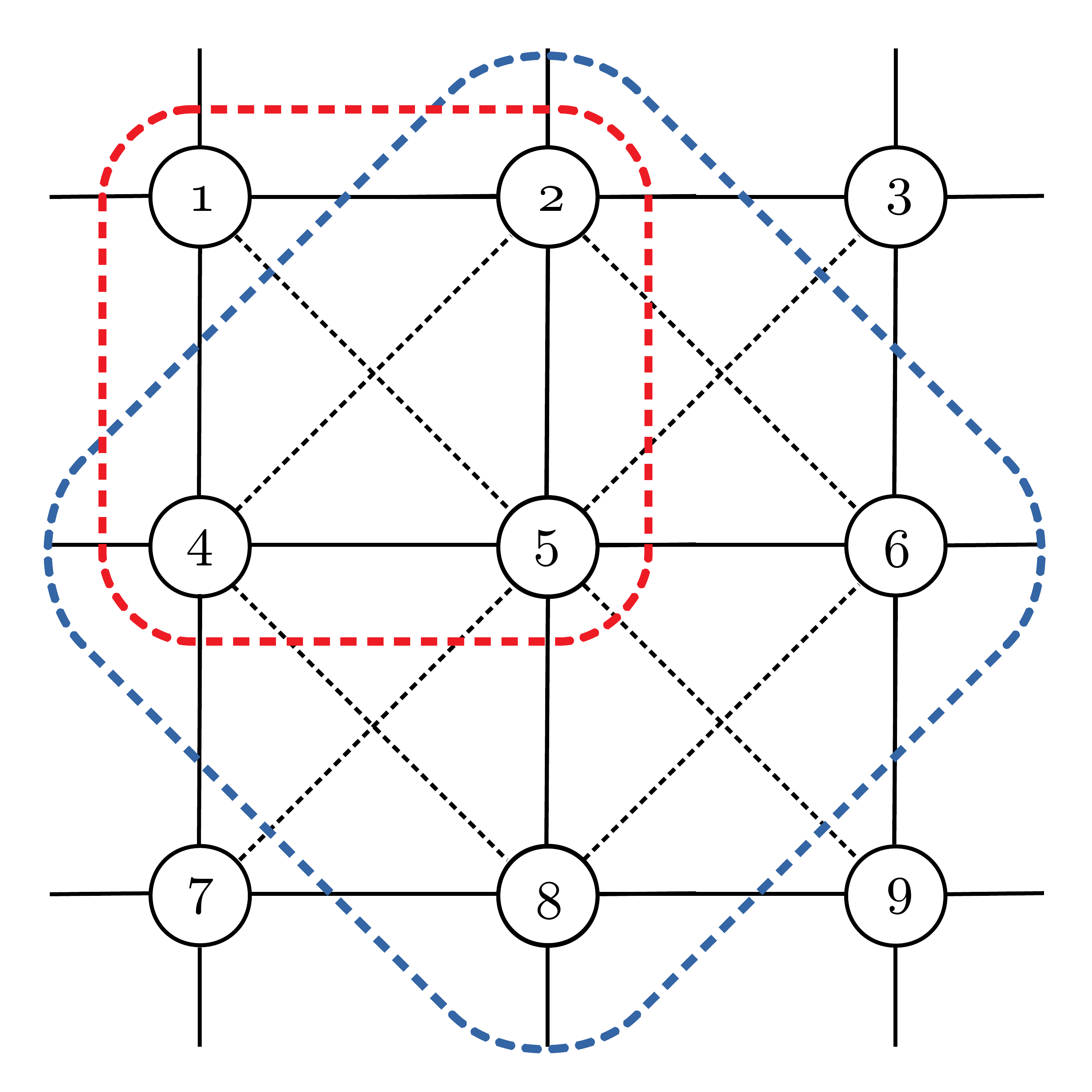}
\caption{The blue dashed rhombus represents the alternative basic plaquette that generates diagonal pairwise regions and effective fields. Every spin and its four nearest neighbors form such a plaquette. Notice that the overlap of contiguous rhombic plaquettes can be a NN pair or a NNN one. In red, the standard (used here) plaquette region.}
\label{fig:rhombic_kikuchi}
\end{figure}

Mean field treatments usually replace real interactions by the interaction with the mean value of certain variables. For example, terms like $J_1 \pauz_1 \pauz_2$ might be roughly approximated by $J_1 \pauz_1 \<\pauz_2\>$
or something similar, which looks like an effective field acting on $\pauz_1$. QCVM effective fields can be understood in a similar way. With this in mind,
we would like to point to what might be a caveat in the choice of regions used in the Kikuchi approximation. Since the overlap of elementary plaquettes corresponds only to NN pairs, the effective fields will only
appear for those interactions. Notice for example the scheme for the single
site distribution $\r_s$ in Figure (\ref{fig:spin_fields}), where no diagonal effective fields are present. It would be desirable to have 
a setup where consistency is also forced for diagonal distributions
which would then appear as diagonal effective fields. We believe this could
be achieved if the elementary plaquettes are defined with a rhombic shape as shown in Figure (\ref{fig:rhombic_kikuchi}).

\section{Results and Discussion}
\subsection{Summary of the classical model}

We start this section presenting the phase diagram of the classical version of the model derived using the Cluster Variational Method\cite{Ours}. On one hand, it may help to understand better the effects of quantum fluctuations, on the other it serves as a checking point to probe that the approximation reproduces the known phenomenology of the model.

\begin{figure}[!htb]
\centering
\includegraphics[width=.4\textwidth]{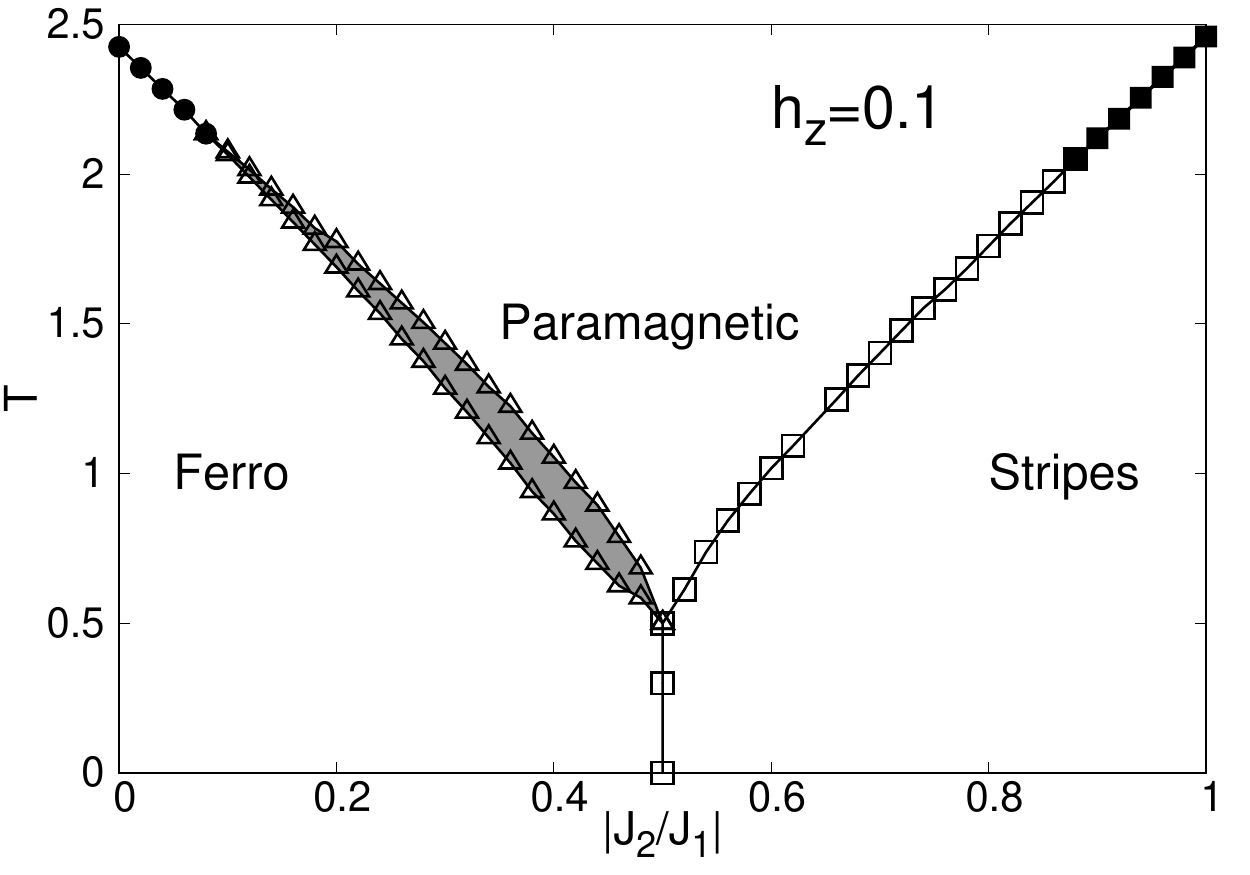}%phase_diagram_Ferro_J1J2.png}
\caption{$T$ vs. $g$ phase diagram in the zero field scenario for the $J_1$-$J_2$ frustrated ferromagnetic Ising model.  Open squares represent discontinuous transitions and filled squares continuous transitions. Triangles represents points in the edge of a non convergence region. }
\label{Zero_Field_Ferro}
\end{figure}

In short, as is shown in Fig.\ref{Zero_Field_Ferro}, depending of the Temperature and the ratio $|J_2|/J_1$ the model may be in one of three phases, ferromagnetic, stripes or paramagnetic \cite{Jin2012}. The lines are guides to the eyes, and the symbols represent the predicted order of the transition, continuous (closed), discontinuous (open).

The ground state in the low $|J_2|$ regime is Ferromagnetic, and the systems behaves essentially as an Ising ferromagnet with a continuous phase transition between the ferromagnetic and the paramagnetic phases. When $J_2=0$ the critical temperature predicted by the approximation is around $T^{*}_c=2.425$, which is close to the exact critical value $T_c=2.26$. In this branch of the phase diagram we notice also a grey zone separating the ferromagnetic and the paramagnetic phases for $0.15 \leq g \leq 0.5$. This is a zone of non-convergence of the algorithm associated with oscillations due to the existence of the two symmetric solutions of the ferromagnetic phase. This zone will appear also in the presence of quantum fluctuations and is discussed in more detail in the Appendix A.

\begin{figure}[!htb]
\centering
\includegraphics[width=.35\textwidth]{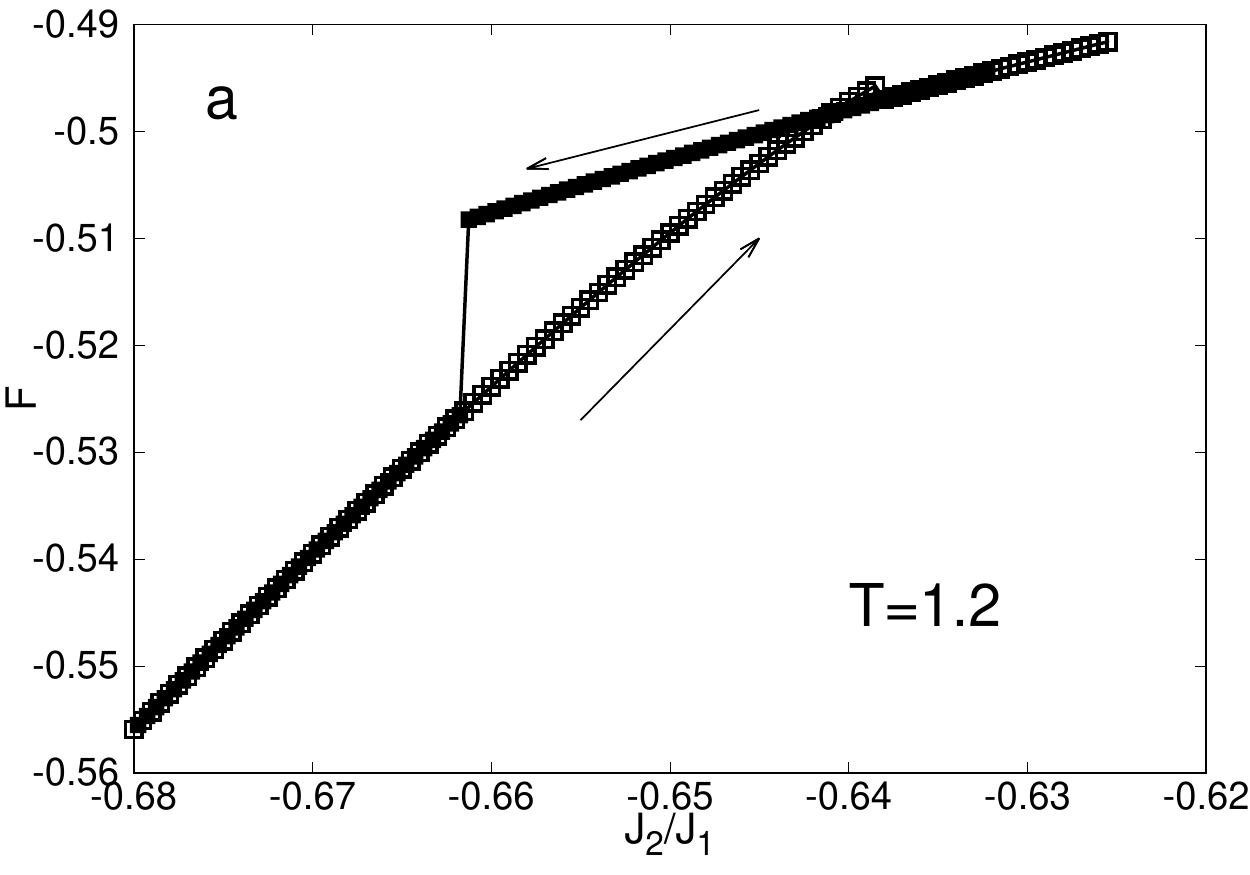}
\includegraphics[width=.35\textwidth]{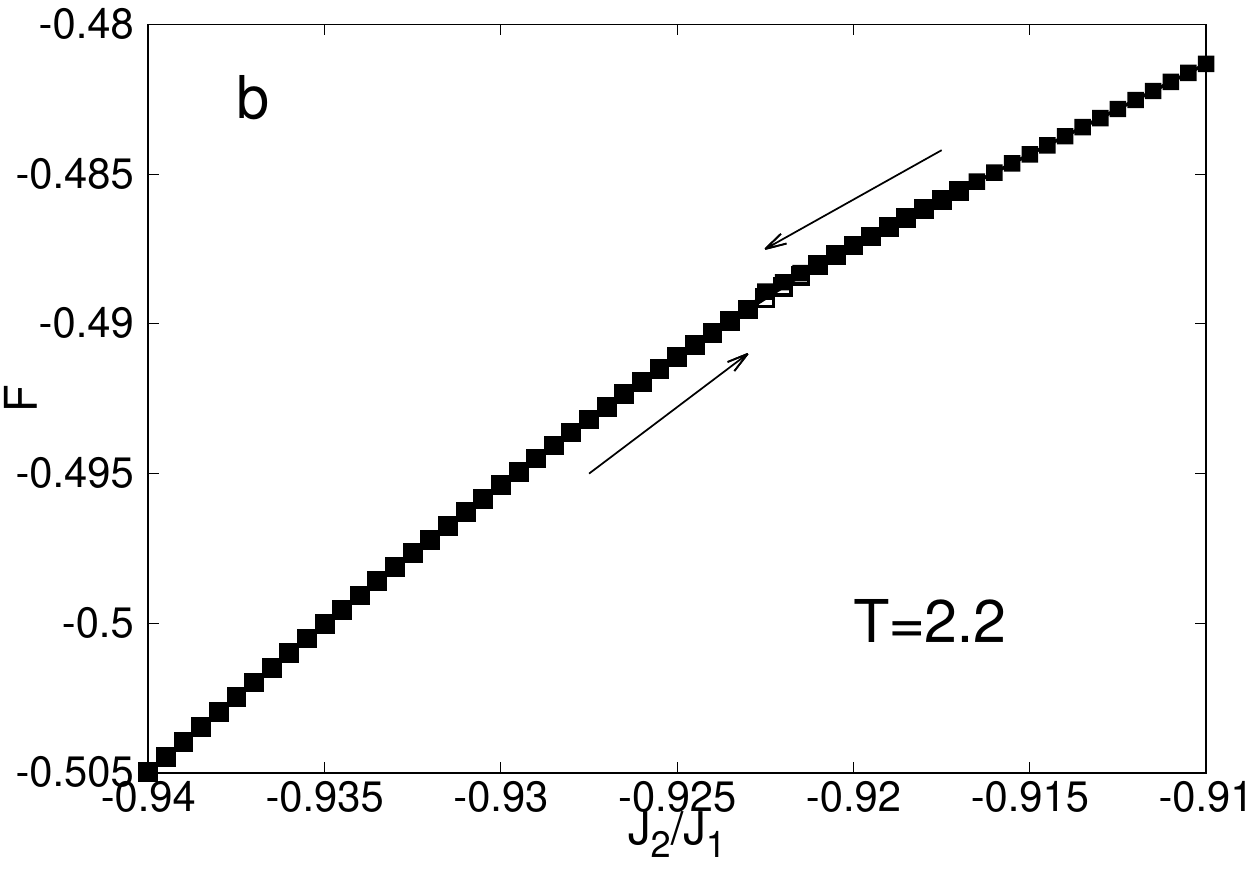}
 \caption{Free Energy curve for two different temperatures, close to the phase transition. In the panel {\bf a}, $T=1.2$, a hysteresis loop is observed. In panel {\bf b}, a well behaved line is observed for $T=2.2$, close to the transition, which occurs around $g=-0.920$.}
\label{Free_Energy_a_b}
\end{figure}

On the other branch of the diagram $g = \frac{|J_2|}{J_1} \geq 0.5$ the system shows a high temperature paramagnetic phase and a low temperature phase of stripes. However the line dividing the two phases is characterized by different kind of phase transitions. Notice  that Jin et al,  \cite{Jin2012} predicted that below $g_c=0.67$ the transition is discontinuous and above continuous. QCVM provides a similar picture, but with a larger $g_c=0.86$. Nevertheless we need to remark that the definition of this critical point from numerical simulations is highly non trivial. For example, in Fig.\ref{Free_Energy_a_b}, we show the behavior of the free energy when the parameter $g$ changes. At $T=1.2$ (Fig.\ref{Free_Energy_a_b}{\bf a}), we find a very clear hysteresis loop,  a signature of a first order transition. On the other hand, for $T=2.2$ (Fig.\ref{Free_Energy_a_b}{\bf b}), the free energy dependence is flat, and there is not evidence of any hysteresis. In the rest of the work the order of the phase transitions was determined both by considering the occurrence of hysteresis loops in the free energy and by making an interpolation of the Free Energy and studying the continuity of its derivative close to the transition as discussed in the previous sections. Results from both methods are equivalent.  %However, there is a rather small prevalence of the paramagnetic state, which is always a solution to the set of Fixed Point equations defined by the QCVM.

\begin{figure}[!htb]
\centering
\includegraphics[width=.35\textwidth]{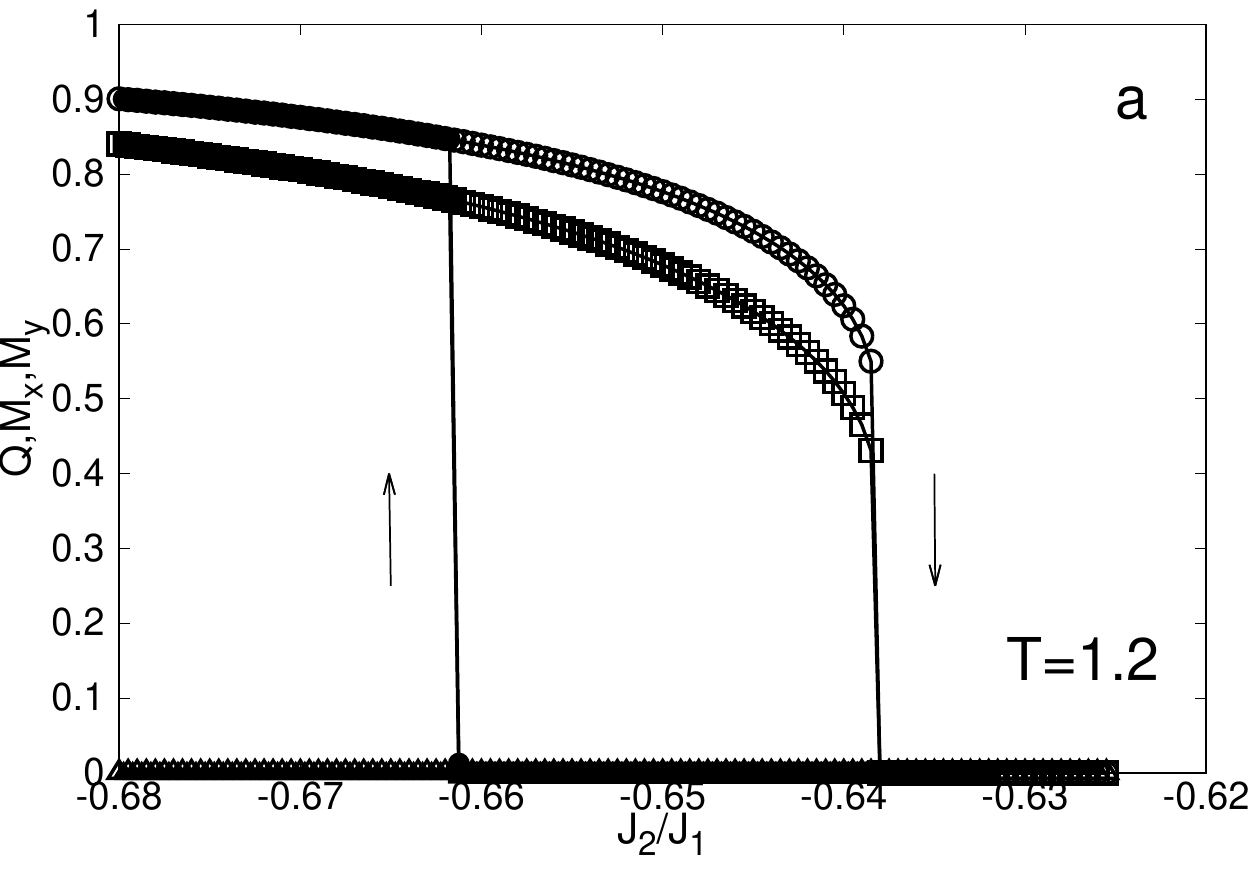}
\includegraphics[width=.35\textwidth]{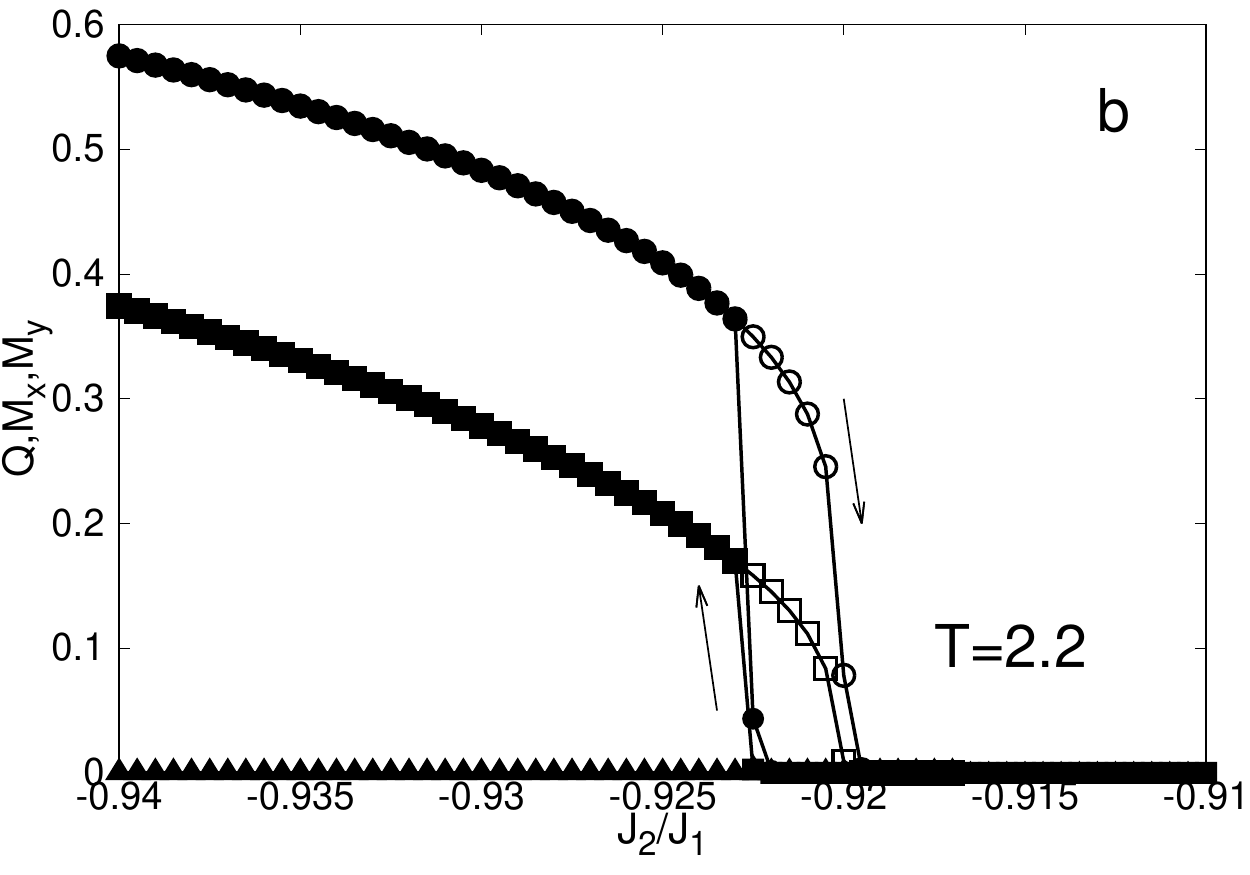}
\caption{Behavior of the order parameters at $T=1.2$ (panel {\bf a}), and $T=2.2$ (panel {\bf b}), close to the phase transition ($g^{*}=0.641$ and $g^{*}=0.920$ respectively). Squares represent the orientational order parameter. Circles and Triangles stand for the Positional order parameters. There are two of them because stripes can form either along the y axes or the x axis. }
\label{order_parameters_diff_Tclassical}
\end{figure}

In Fig.\ref{order_parameters_diff_Tclassical} we  show the behavior of the orientational and positional order parameters around the transition. For  $T=1.2$ (Fig.\ref{order_parameters_diff_Tclassical}{\bf a}), it can be observed a wide hysteresis loop when $J_2$ increases or decreases and a sharp jump in the order parameter, both indications of a discontinuous transition. On the other hand, (Fig.\ref{order_parameters_diff_Tclassical}{\bf b}) for $T=2.2$ there is a continuous change in the order parameter when $J_2$ increases, suggesting a continuous transition. However, there is also a small hysteresis loop suggesting a possible discontinuous transition. We checked that it is rather a consequence of the stability of the paramagnetic solution for this approximation, which is always a valid solution of the fixed point equations.

\begin{figure}[!htb]
\centering
\includegraphics[width=.4\textwidth]{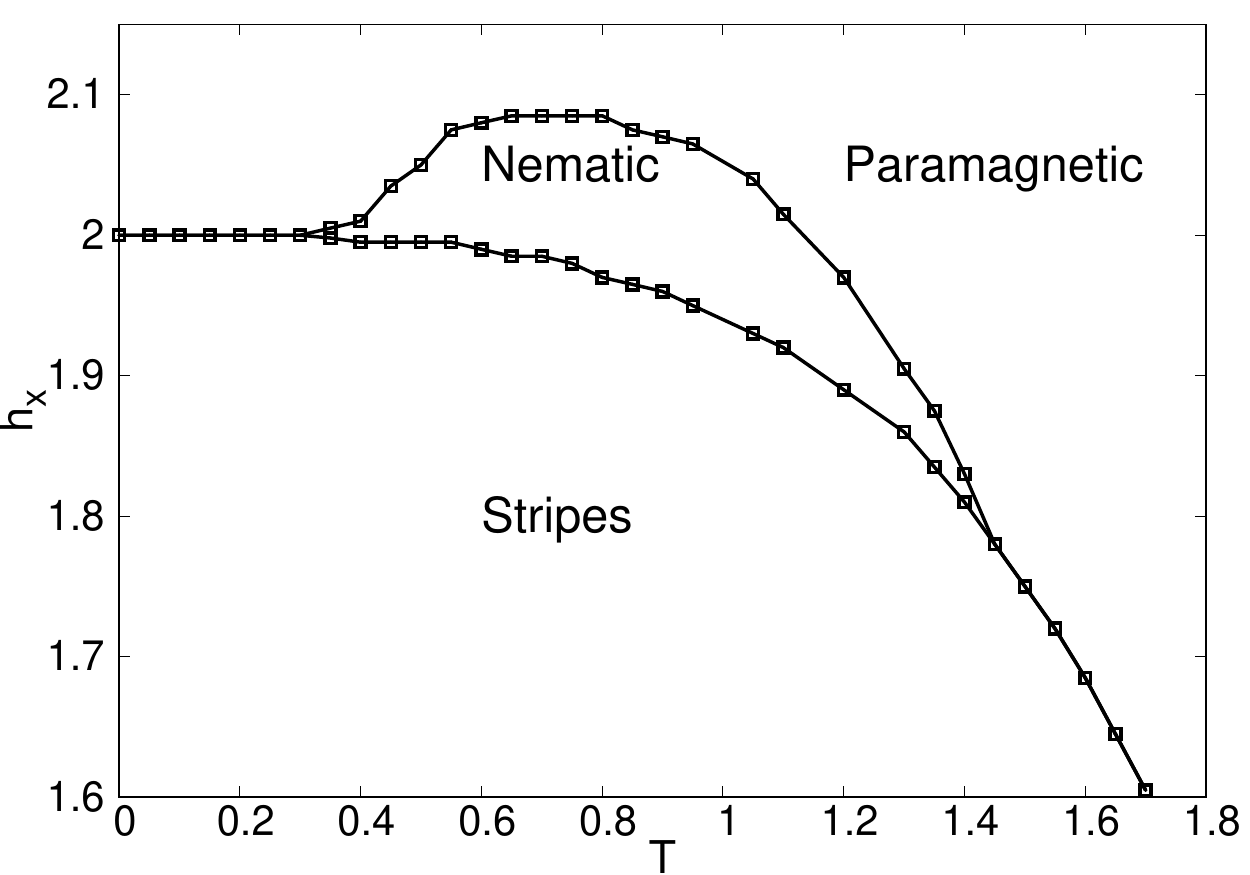}%phase_diagram_Ferro_J1J2.png}
\caption{$h_x$ vs. $T$ phase diagram phase diagram for the classical $J_1$-$J_2$ frustrated ferromagnetic Ising model with $J_1=1$ and $J_2=-1$. Nematic phase is present in a narrow region of the phase diagram for large field, and relatively low values of temperature.}
\label{phase_diagram_GBP}
\end{figure}

The richness of this model becomes more evident in the presence of a longitudinal field. In this case (see Figure \ref{phase_diagram_GBP}) for proper values of the temperature and the field a nematic phase separates the paramagnetic phase and the phase of stripes \cite{Stariolo, Ours}.

Summarizing, in the absence of external fields the classical $J_1$-$J_2$ model presents (depending on the value of $J_2$) two low temperature phases, ferromagnetic and stripes. Increasing $T$ the ferromagnet behaves essentially like an Ising ferromagnet with a continuous transition to a paramagnetic phase. The phase of stripes may have, depending on $J_2$  a continuous or discontinuous transition to the paramagnetic phase. On the other hand, and depending on the temperature, in the presence of a longitudinal field the stripe and nematic phases may be separated by a novel nematic phase.

\subsection{The role of quantum fluctuations}

With the previous understanding of the classical model we now concentrate our attention in the main motivation of our work: the role of quantum fluctuations.

\begin{figure}[!htb]
\centering
\includegraphics[width=.4\textwidth]{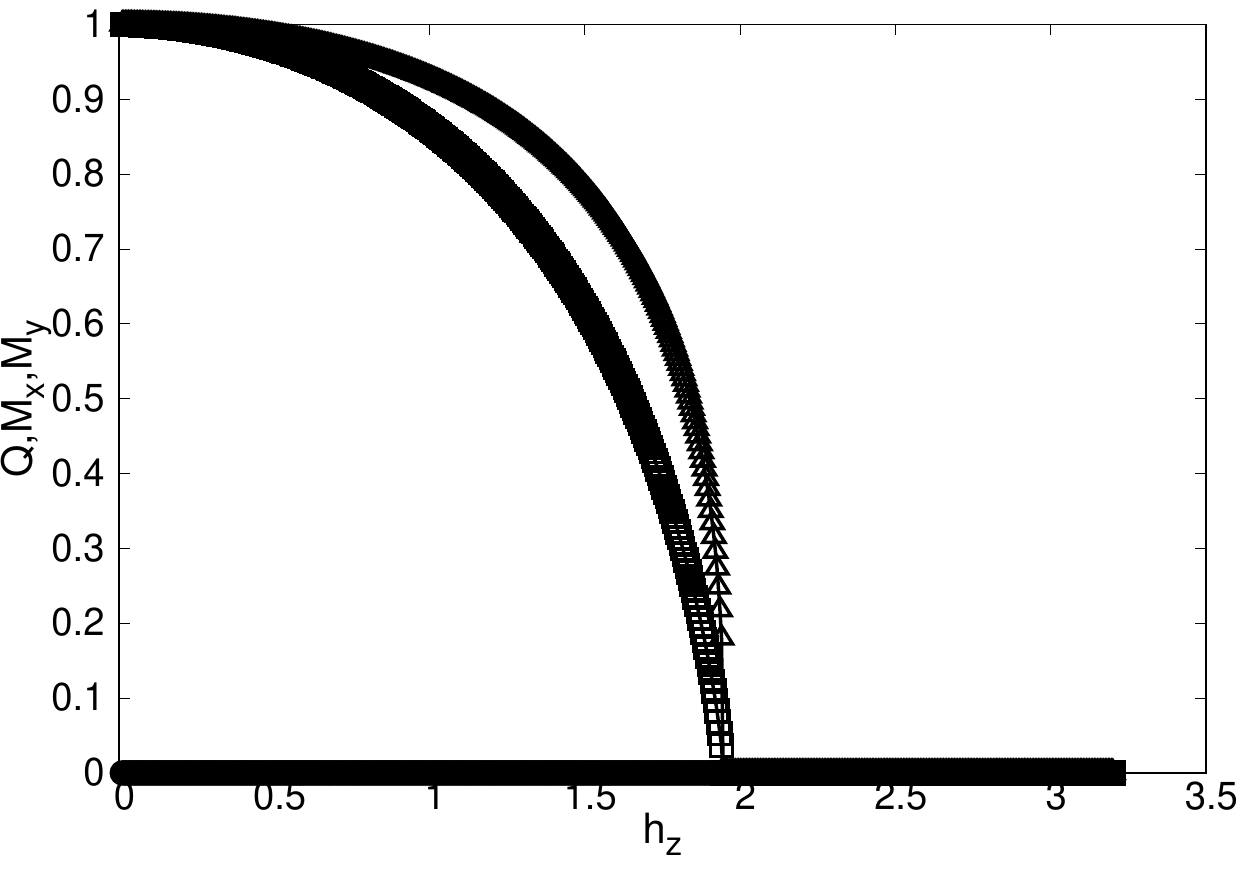}%phase_diagram_Ferro_J1J2.png}
\caption{Behavior of the orientational and positional order parameters for $T=0.1$, $h_x=0$, $J_2=-0.72$, close to a quantum phase transition induced by a transverse field. The critical value of $h_z$ is $h^{*}_z=1.945$.}% Symbol convention is the same as in Fig.\ref{order_parameters_diff_T}.}
\label{order_parameters_hz}
\end{figure}

We first explore the behavior of the QCVM at low values of the temperature, $T=0.1$, and in the absence of the longitudinal field, $h_x=0$. In this regime, and in the absence of quantum fluctuations the system is either in a ferromagnetic or in a stripe phase (see Fig. \ref{Zero_Field_Ferro}). Once the transverse external field is turned on and increased the system eventually turns paramagnetic with both, the orientational and positional order falling to zero as is shown in Fig.\ref{order_parameters_hz}. 

\begin{figure}[!htb]
\centering
\includegraphics[width=.4\textwidth]{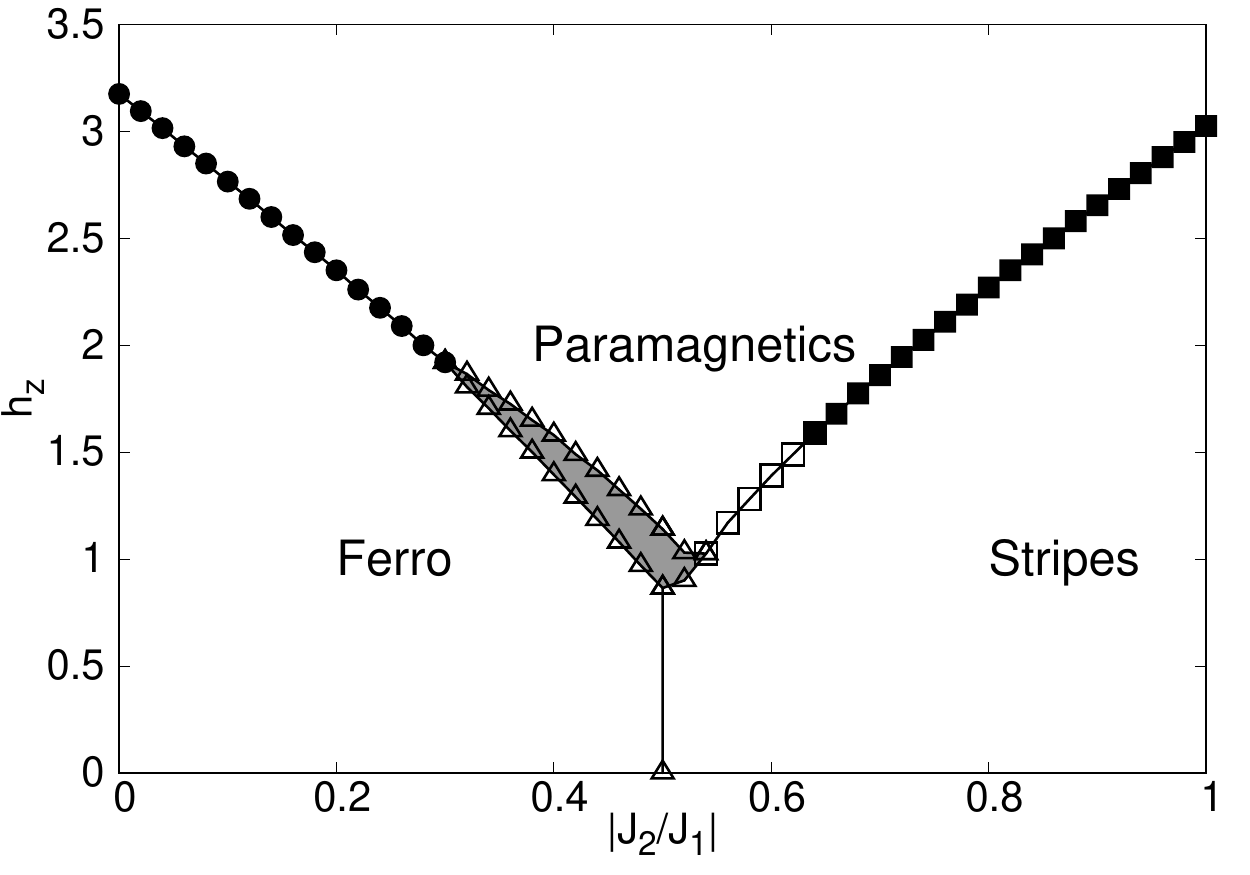}
\caption{$h_z$ vs $g$ phase diagram for $T=0.1$  for the $J_1$-$J_2$ frustrated ferromagnetic Ising model. In the stripes-paramagnetic transition a change in the nature of the phase transition occurs around $g^{*}_c=0.64$, $h_z=1.59$. The election of the symbols follows the same convention as in Fig.\ref{Zero_Field_Ferro}. }
\label{Ferro_hz}
\end{figure}

An extensive study of the parameters of the model at low temperatures allows the construction of the $h_z-g$ phase diagram (see Figure \ref{Ferro_hz}). Also in this case the ground state for $g \leq 0.5$ is a ferromagnet, and the phase transition is continuous. Moreover, the predicted critical field at $J_2=0$, corresponding to the quantum Ferromagnetic Ising Model is around $h^{*}_z=3.175$\cite{ourQCVM}, which is close to the exact value $h_{z(c)}=3.04$\cite{MonteCarloIsing}.
In the right branch of the diagram,  the critical point, where the discontinuous transition between the phase of stripes and the paramagnetic phase  becomes continuous is around $g^{*}_c \approx 0.64$, with $h^{*}_z=1.590$. This value is also similar to the one reported by means of a Cluster Mean Field approach in \cite{Kellerman2019} ($g_c=0.56$). On the other hand, QCVM for $J_2=-1$ predicts a critical transverse field $h^{*}_{z}=3.025$, much lower than the one reported in \cite{Kellerman2019}, $h_z=3.658$.

\begin{figure}[!htb]
\centering
\includegraphics[width=.4\textwidth]{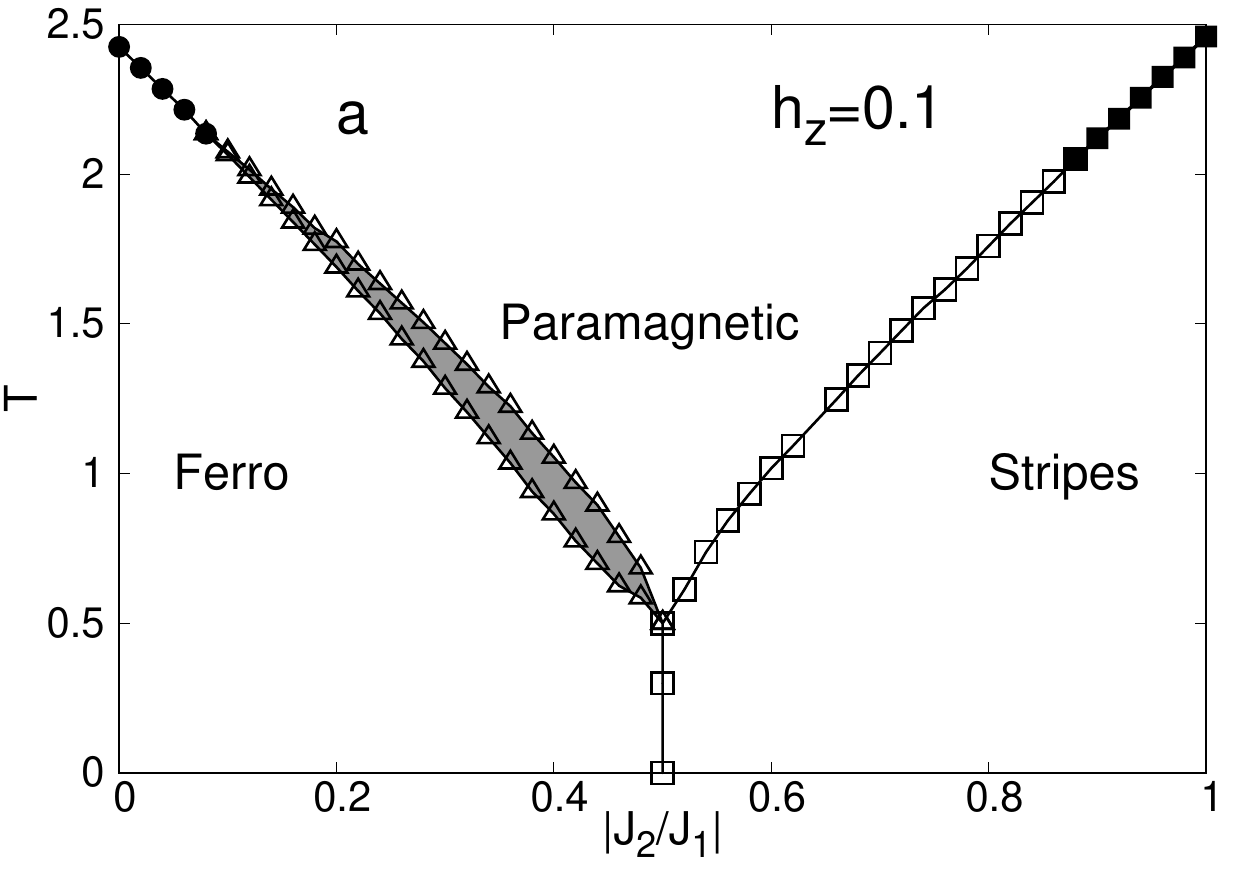}
\includegraphics[width=.4\textwidth]{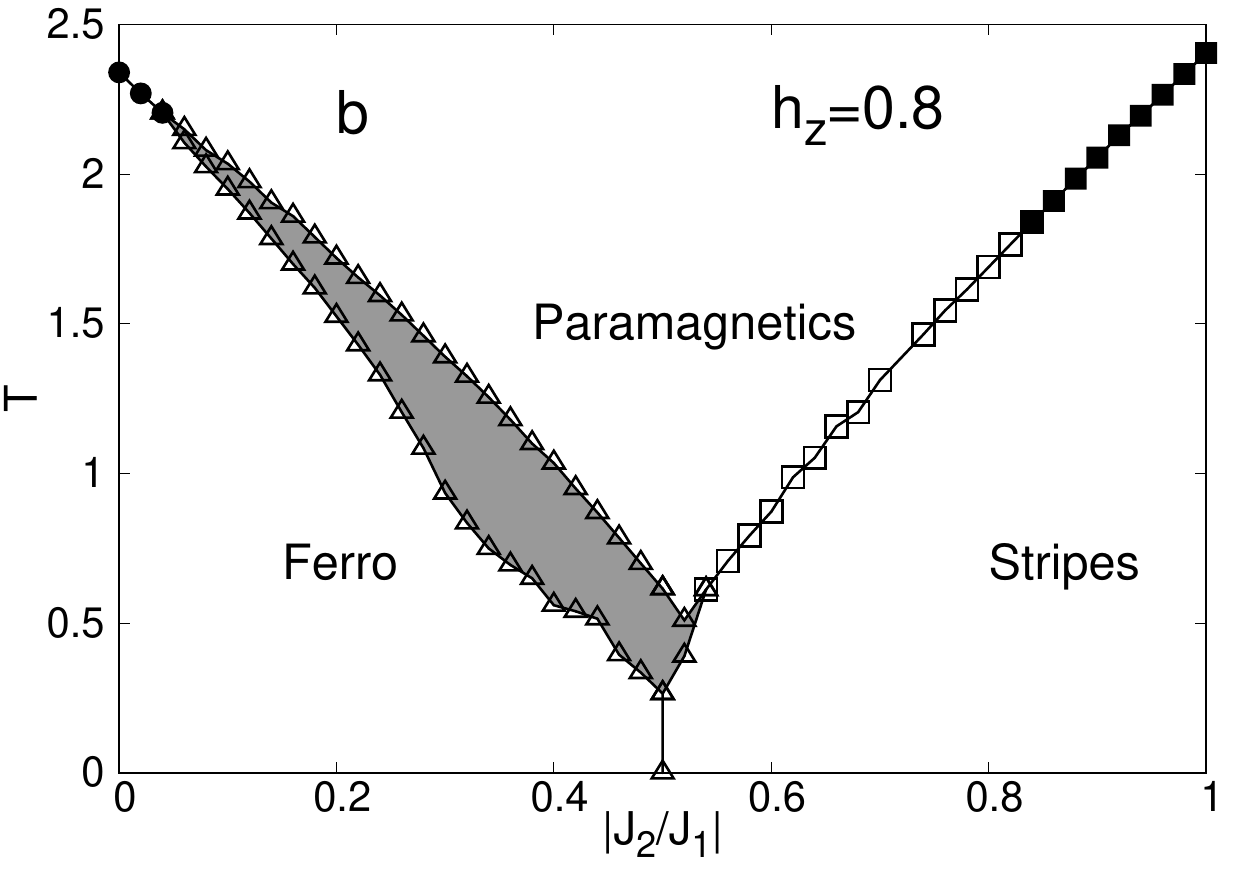}\\
\includegraphics[width=.4\textwidth]{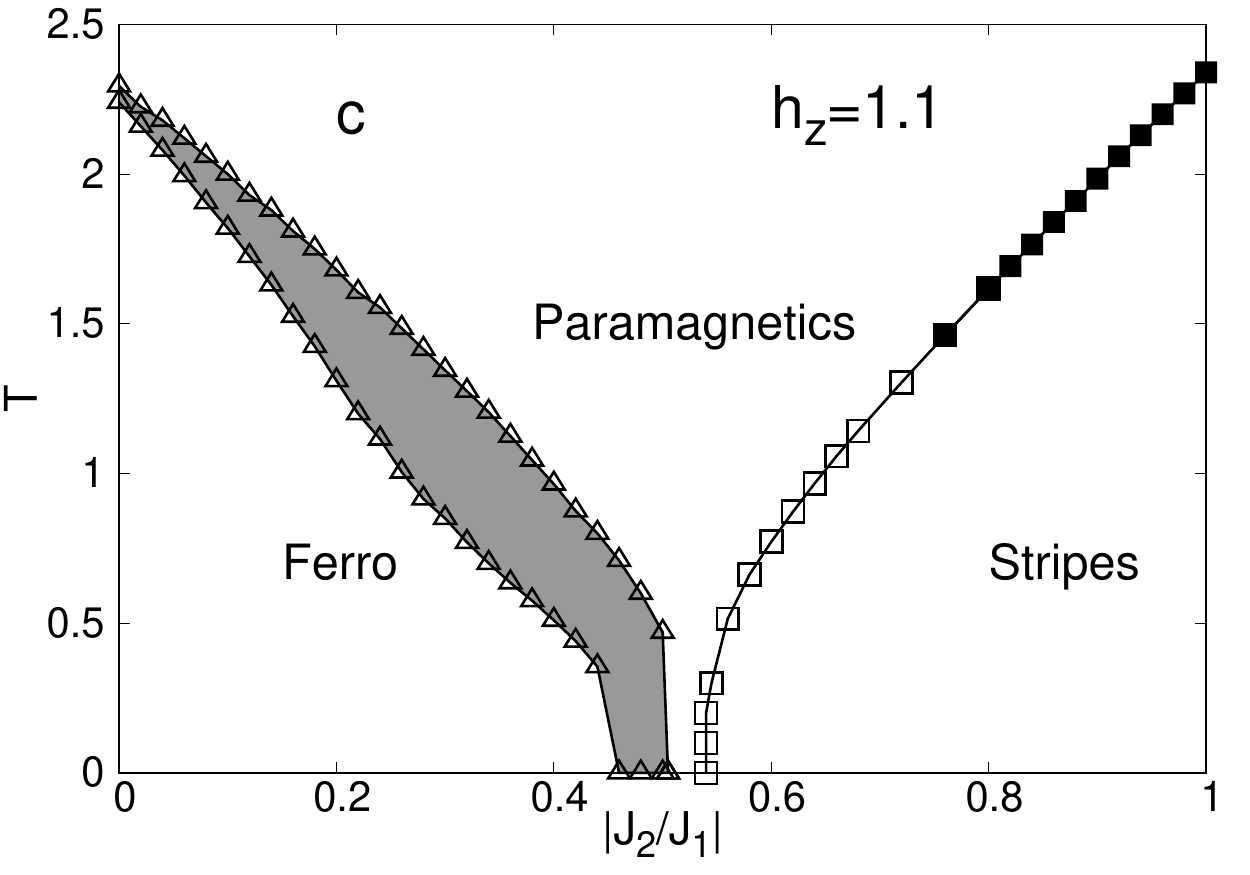}
\includegraphics[width=.4\textwidth]{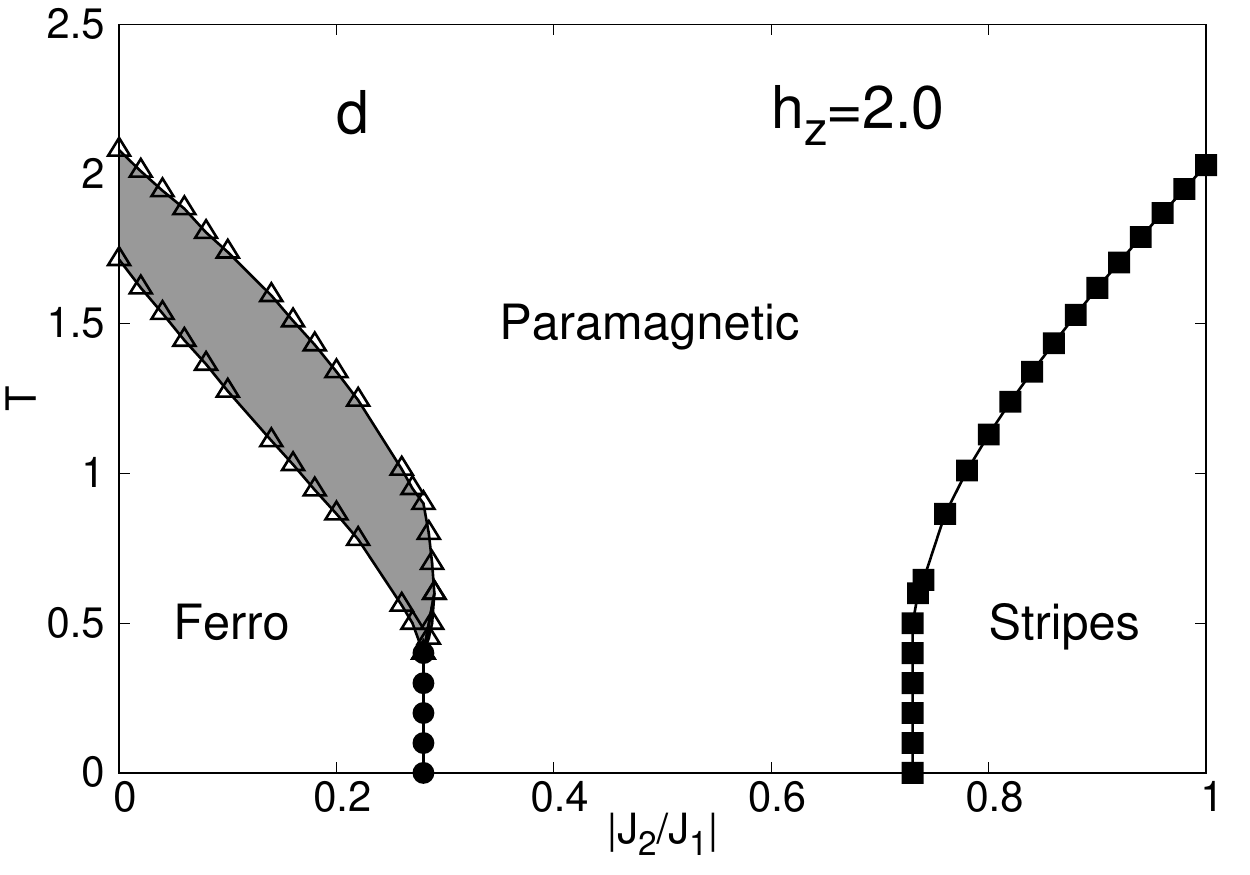}
\caption{$g$ vs. $T$ phase diagrams in the presence of transverse field. Only for large values of $h_z$ do differences in the phase diagram become relevant. A gap paramagnetic region appears between ordered phase at zero temperature for larger transverse fields.  }
\label{effect_hz_T}
\end{figure}

The effect of quantum fluctuations reshapes the classical phase diagram $T-g$ presented in Figure \ref{Zero_Field_Ferro}. This is  shown in Fig. \ref{effect_hz_T}. At low values of $h_z$, quantum fluctuations just extend the range of values of the ratio $|J_2|/J_1$ at which the paramagnetic to stripe transitions is continuous, i.e., in practice shifting $g_c$ to the left. On the other hand, when the transverse field is large enough ($h_z \geq 1$), quantum fluctuations induce  a gap between the ferromagnetic and the phase of stripes that is ocupied by the paramagnetic phase. In this case, only when the frustration is very small $J_2 \ll J_1$ or when it is very large $J_2 \sim J_1$ the system exhibits actual order at low temperatures.  

\begin{figure}[!htb]
\centering
\includegraphics[width=.4\textwidth]{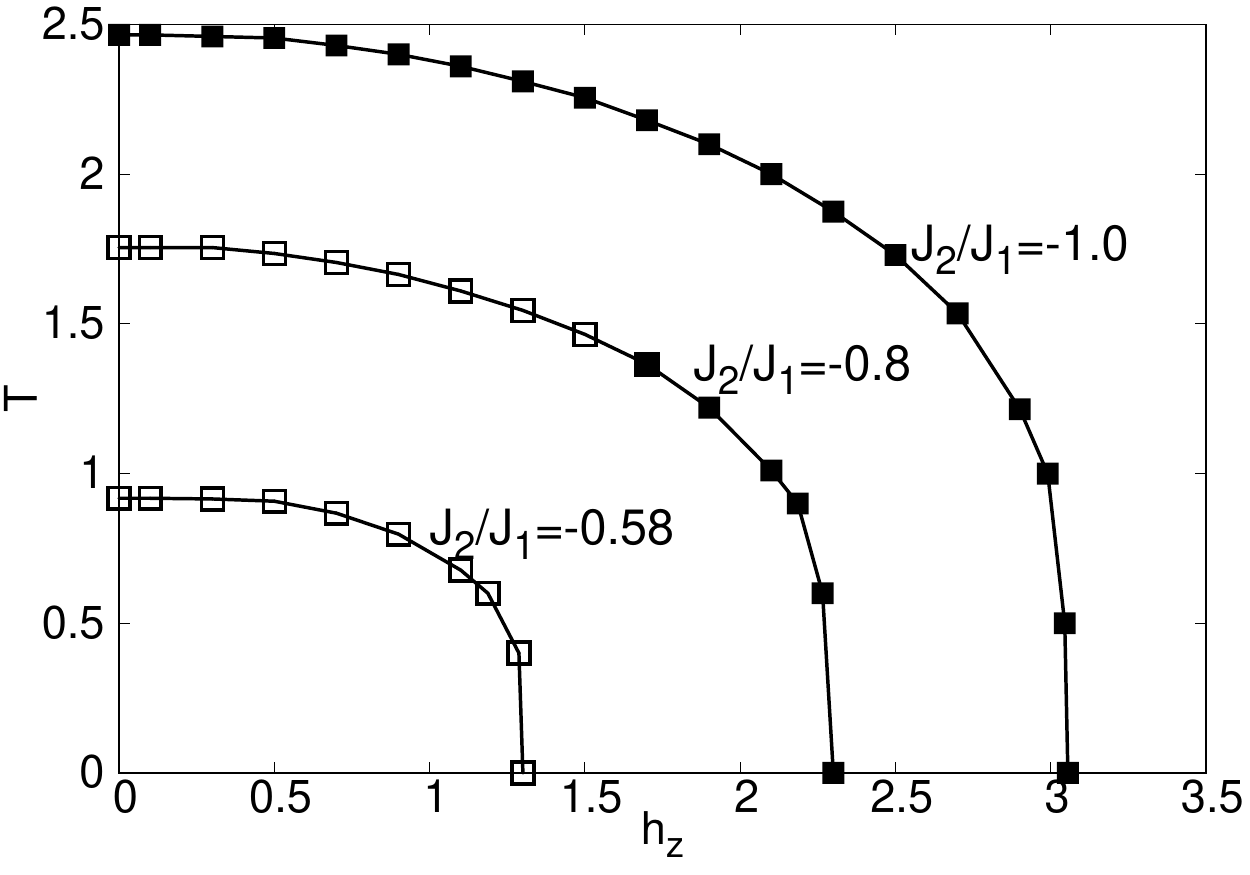}
\caption{$T$ vs $h_z$ phase diagrams for different values of $J_2$. For $J_2=-0.58$, a line of first order transition separates the stripes phase from the paramagnetic one. For $J_2=-1$, a line of second order transition is observed. For $J_2=-0.8$, a line with both first and second order transition is observed, with a critical point located around $h_z=1.7$. First Order transition is represented with open squares, and the continuous transition with filled squares.}
\label{phase_diagram_T_vs_hz}
\end{figure}

To summarize, we show  in Fig.\ref{phase_diagram_T_vs_hz} the $T$ vs $h_z$ phase diagrams for different values of the ratio $J_2/J_1$. Three different scenarios are well characterized. For example, if $J_2/J_1=-0.58$, we find a  full line of discontinuous phase transitions. In the other extreme, when $J_2/J_1=-1$, there is only continuous transitions, while at intermediate values of the frustration (for example, $J_2/J_1=-0.80$), we can observe a line of mixing continuous and discontinuous phase transitions. In other words, in the presence of quantum fluctuations, the frustration favors the occurrence of continuous transitions.

\subsection{Quantum fluctuations in the presence of a longitudinal field}

In this section we study the model in the presence of a longitudinal field, $h_x$. As we already discussed, in the absence of the transverse field, the presence of a longitudinal field induces a nematic phase that separates the paramagnetic and the phase of stripes in a wide range of temperatures (see Figure \ref{phase_diagram_GBP}). We will show that quantum fluctuations make this scenario more plausible with a nematic phase that may penetrate the phase of stripes.

\begin{figure}[!htb]
 \centering
 \includegraphics[width=.4\textwidth]{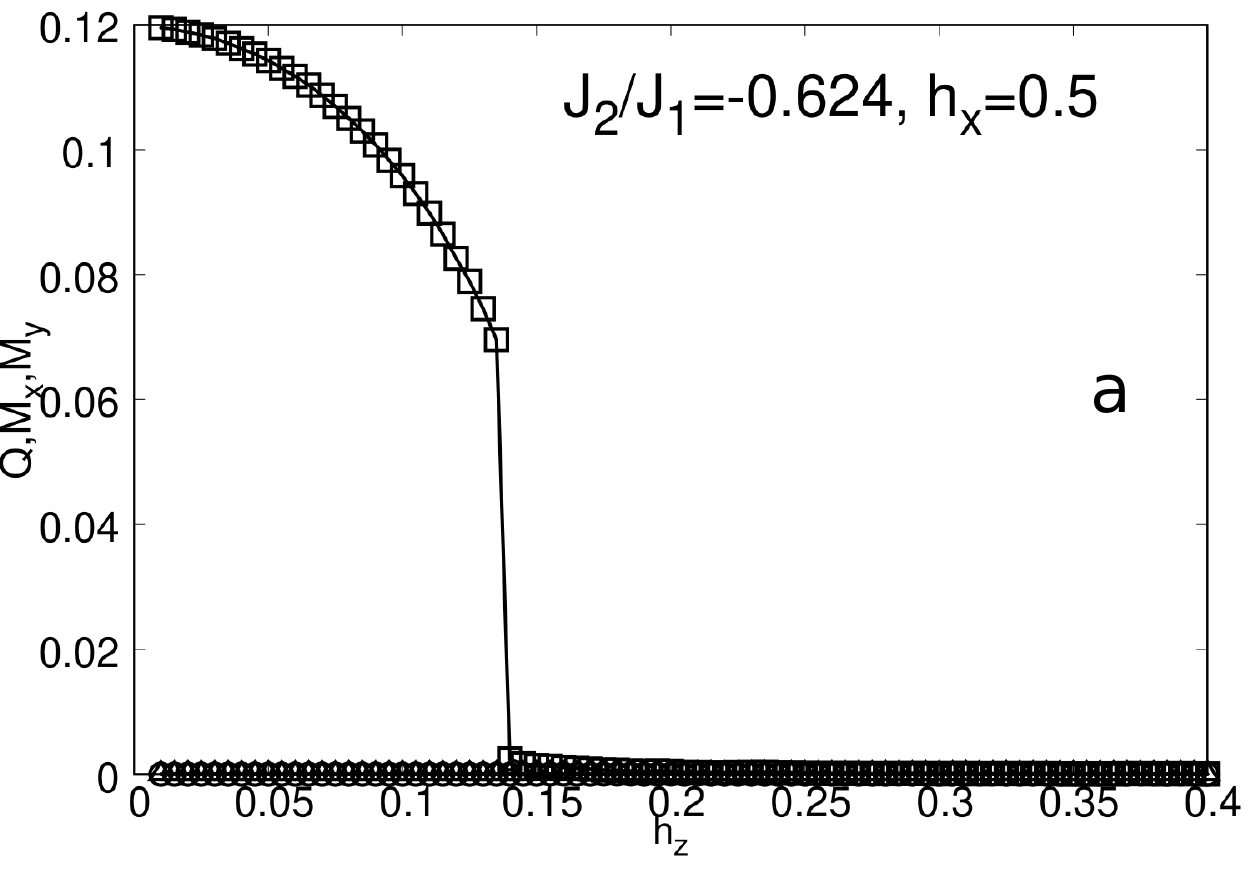}
 \includegraphics[width=.4\textwidth]{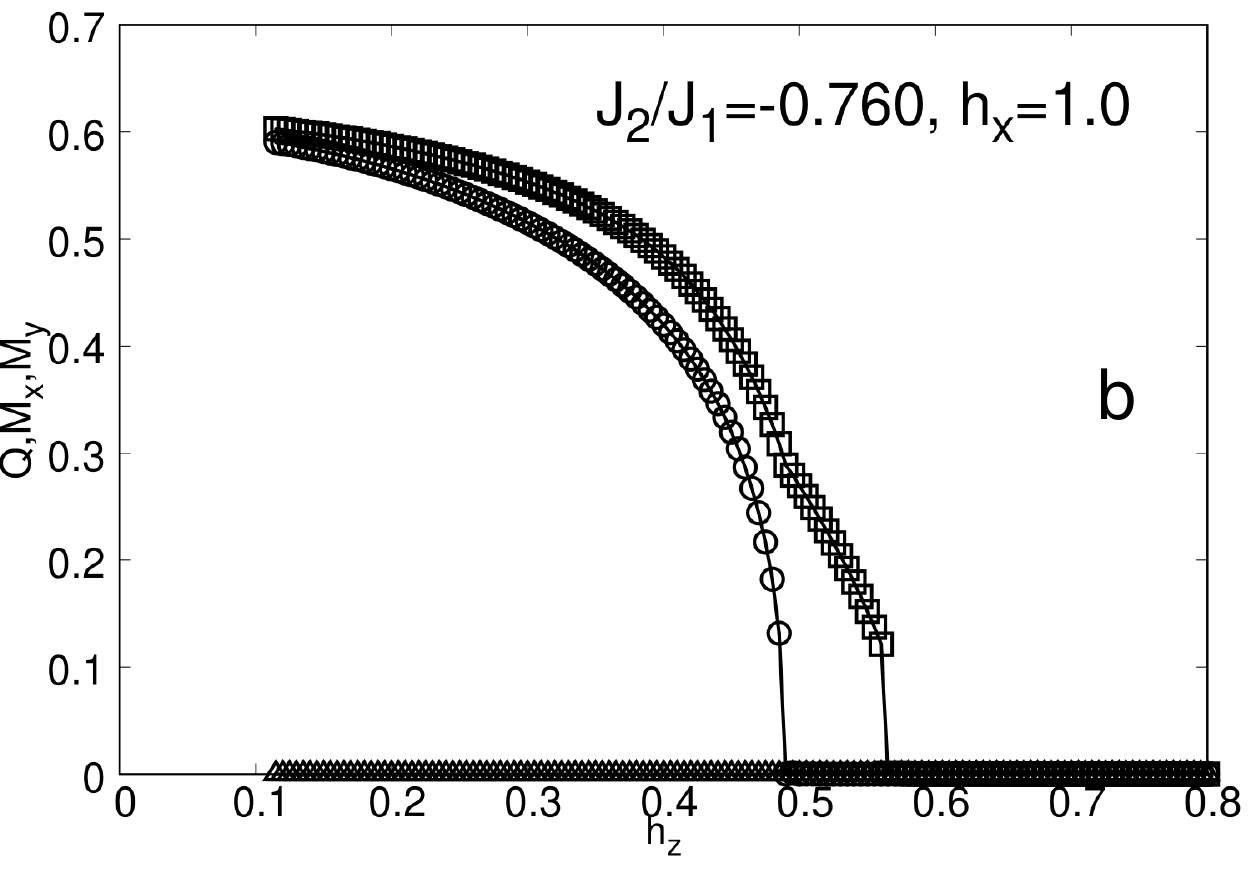}
\caption{Behavior of the orientational and positional order paramaters against the transverse field, at the same temperature for two different values of frustration and longitudinal fields. Panel {\bf a} corresponds to the parameters $J_2/J_1=-0.624$ and $h_x=0.5$. Panel {\bf b} corresponds to $J_2/J_1=-0.76$, $h_x=1.0$. }
\label{effect_hz_Order_Param}
\end{figure}

To have a glance on the effect of the longitudinal field in the picture discussed so far, in Fig.\ref{effect_hz_Order_Param} we show the behavior of the orientational and positional order parameters against the transverse field for different values of the ratio $g=J_2/J_1$ and the field $h_x$. It can be observed that depending on the values of $g$ and $h_x$ the system may well have only orientational order but lack of translation order ($Q>0,M=0$) if $h_z$ is small enough (panel {\bf a}), or alternatively (panel {\bf b}), a phase of stripes for low values of $h_z$ in which both $Q$ and $M_x$ are different from zero. The first is a signature of the nematic phase and a transition from a nematic to a paramagnetic phase due to quantum fluctuations. In the second case, a scenario like that in panel {\bf b} may take place in which, increasing $h_z$, first the magnetizations structure homogenizes, continuously driving the system into a nematic phase, and only later when $h_z \geq 0.56$, the orientational order parameter ($Q$), drops abruptly to zero.

\begin{figure}[!htb]
\centering
\includegraphics[width=.25\textwidth]{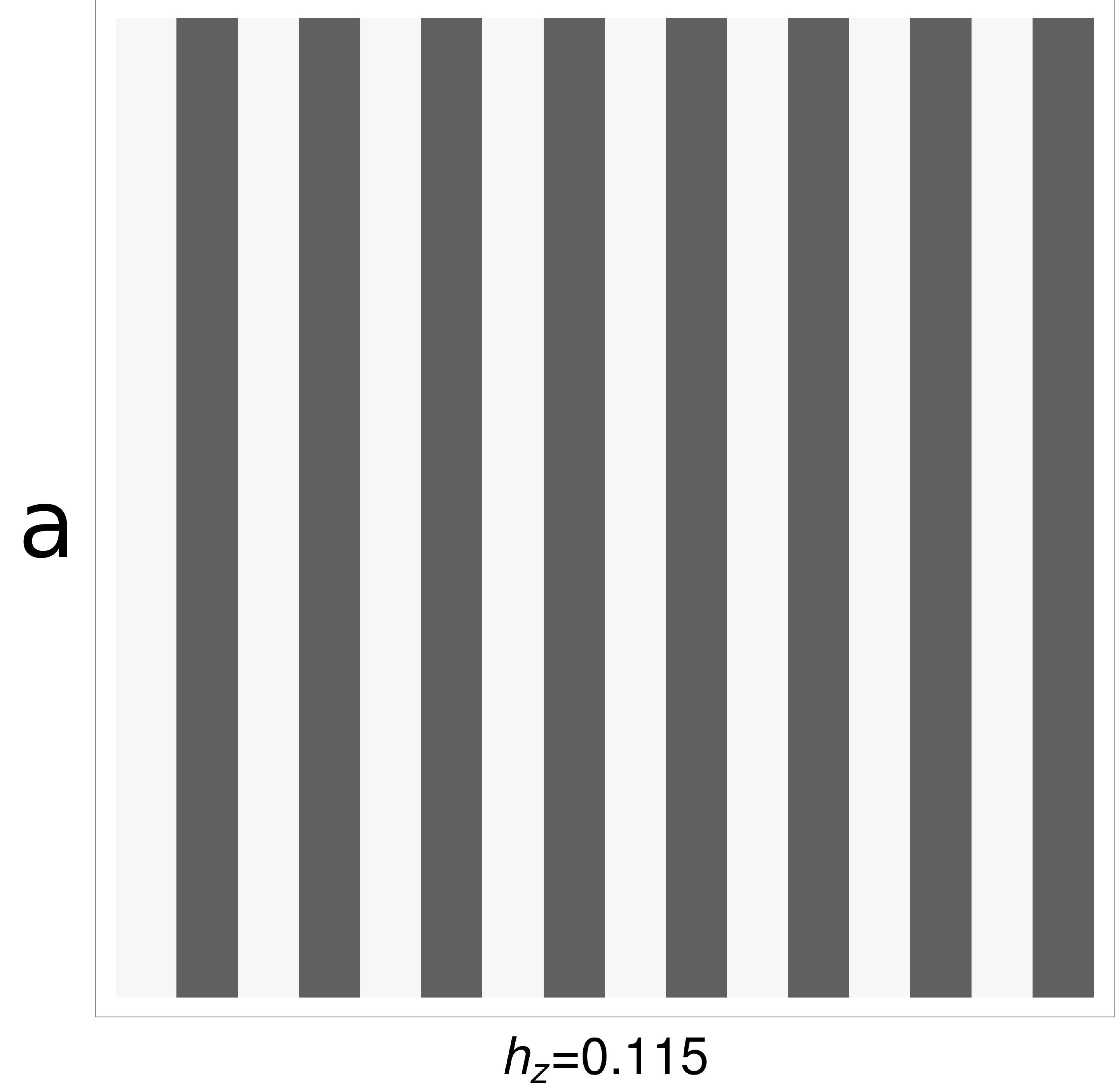}\includegraphics[width=.225\textwidth]{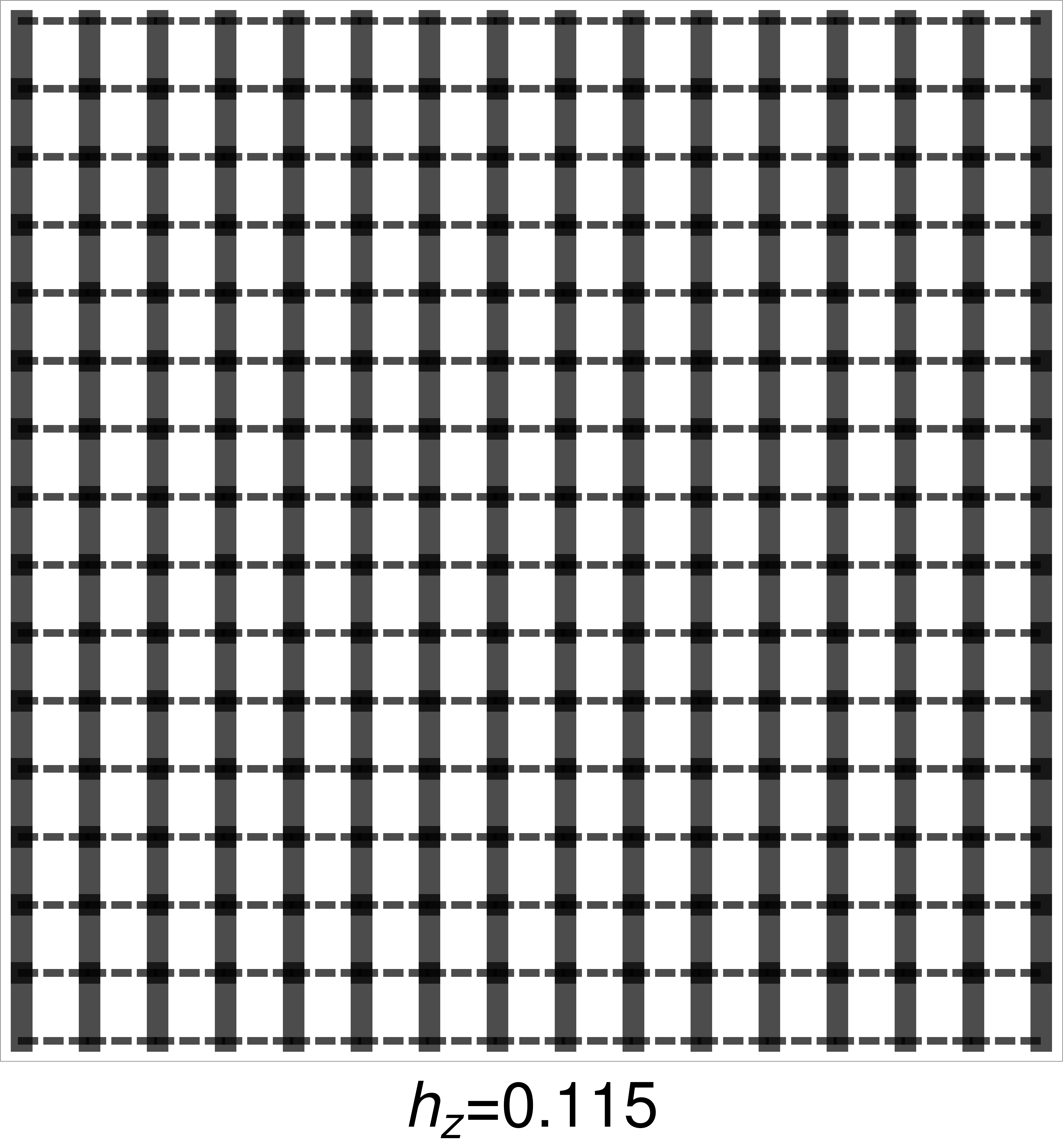}
\includegraphics[width=.25\textwidth]{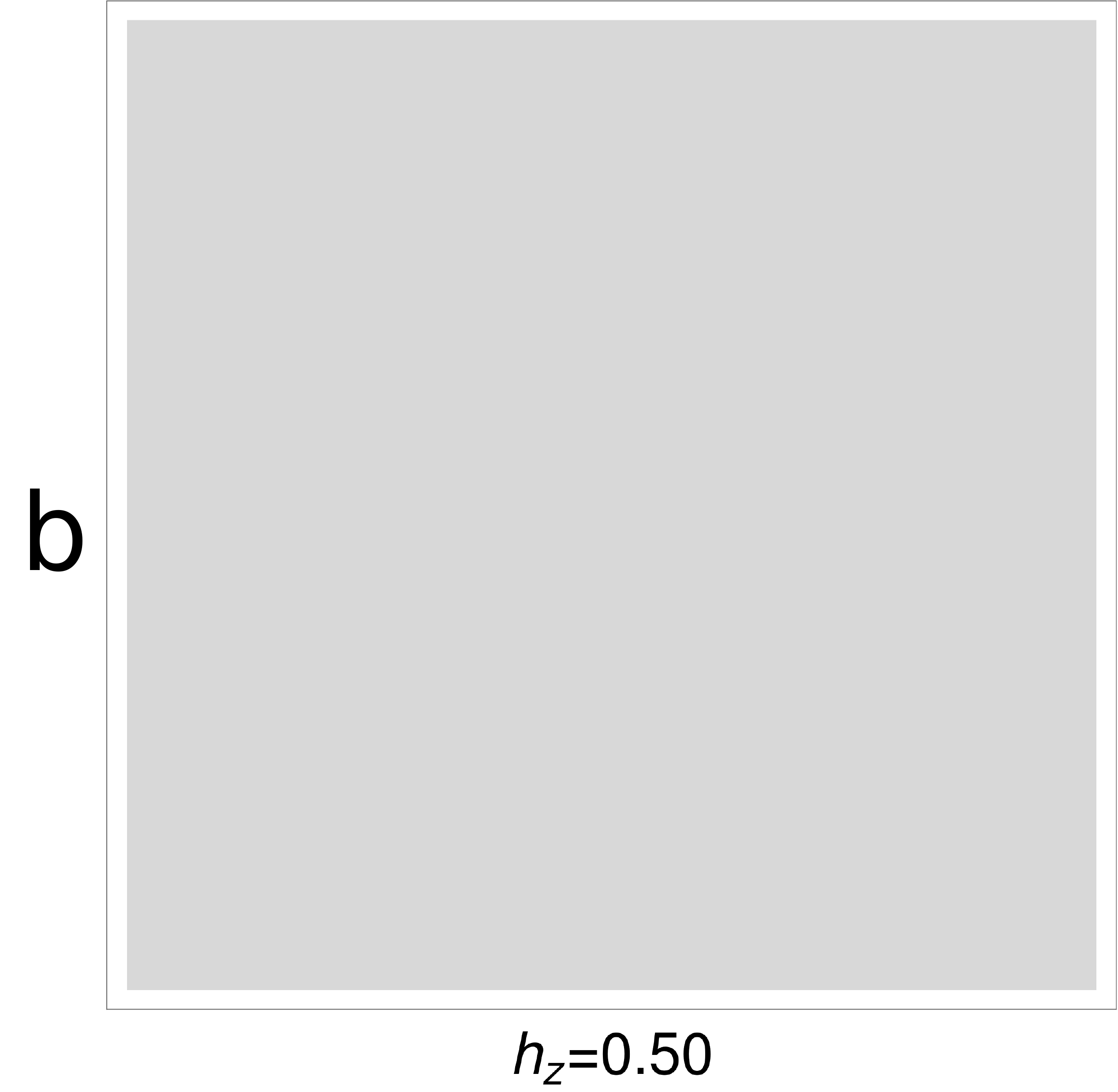}\includegraphics[width=.225\textwidth]{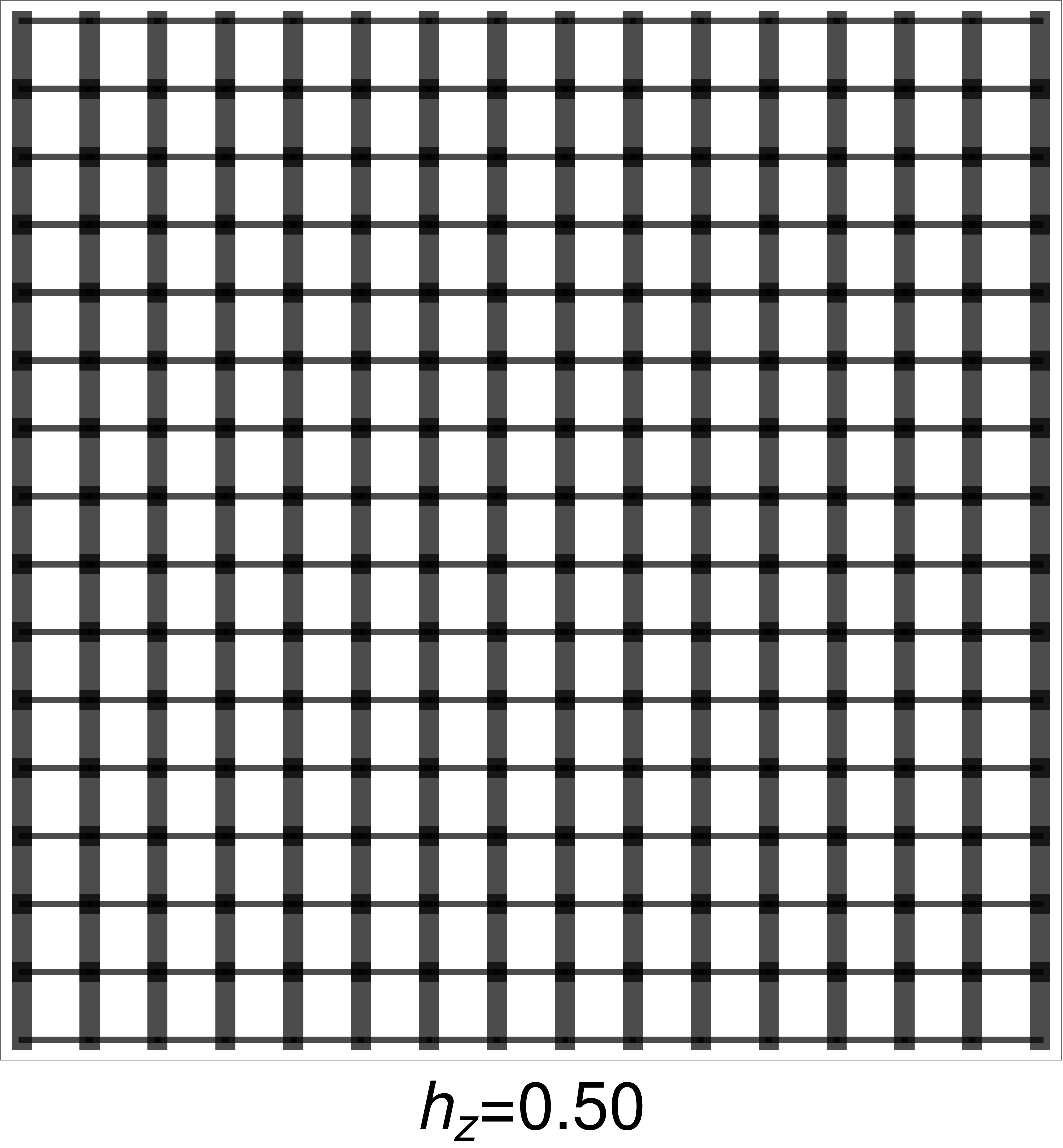}
\includegraphics[width=.25\textwidth]{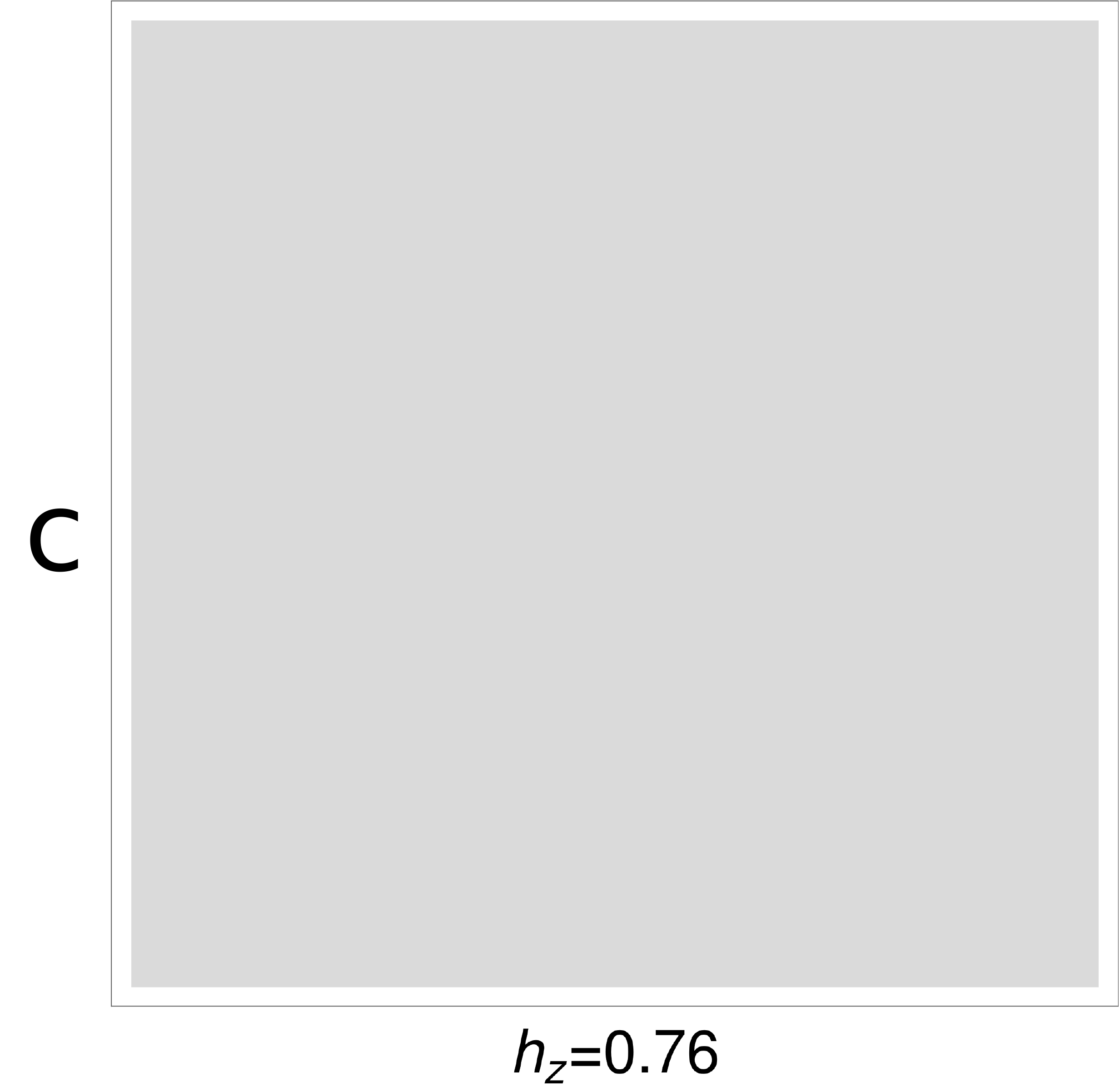}\includegraphics[width=.225\textwidth]{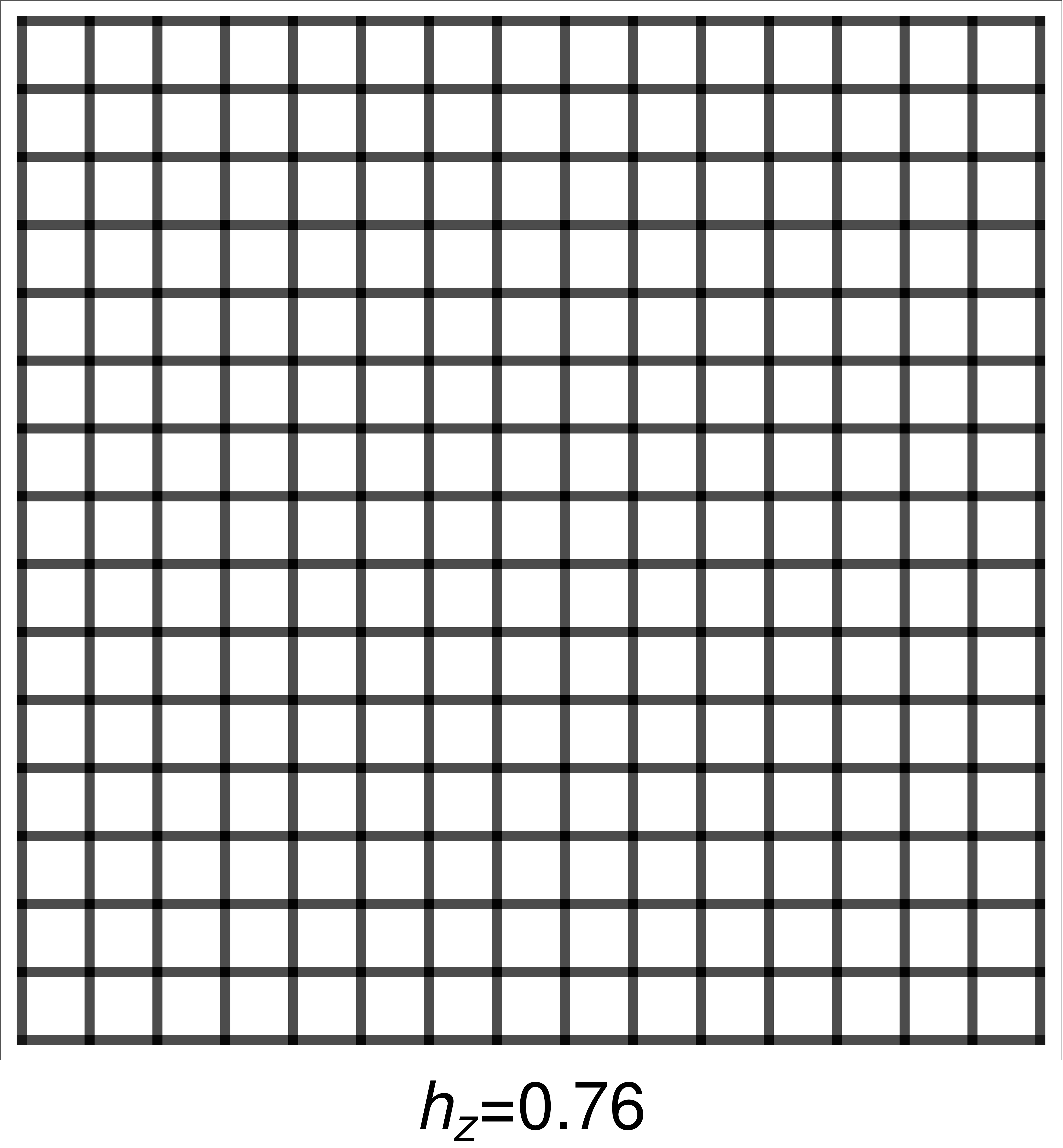}

\caption{Magnetization and correlation structure for different values of $h_z$, with $J_2/J_1=-0.76$, $h_x=1.0$ and $T=0.1$. The width of the lines representing the links is proportional to the module of correlation between the sites it joins. On the other side, negative correlations are represented with dashed lines and positive correlations with continuous lines. The magnetization structure is represented in gray scale so that white corresponds to one and black to -1. }
\label{structure_nem}
%\label{order_parameters_diff_T}
\end{figure}

A clearer picture of the situation appears studying the  evolution of the correlation and magnetization structure in the lattice as in Fig.\ref{structure_nem} for different values of $h_z$. In this picture the correlation structure is represented using solid (dashed) lines to represent positive (negative) correlations, with a width proportional to their modulus. In Fig.\ref{structure_nem}{\bf a}, we can observe a clear structure of stripes, but with the subtle feature that the global magnetization is not zero, as the sites with positive magnetizations (white stripes) have a larger module than those pointing in the opposite direction. This is a consequence of the application of a field in a direction that breaks the symmetry between the two orientations, positive and negative. Another remarkable feature is that the correlation between different stripes (dashed lines) are lower compared to that between elements of the same stripes (solid lines). As the transverse field increases, we approach the nematic phase, represented in Fig.\ref{structure_nem}{\bf b}. There we can observe an homogeneous magnetization, yet a clear remainder of the stripes is observed in the correlation structure. Even when both correlations are positives we can clearly see that the line joining sites of the same column are represented with wider lines than those between different stripes. Lastly, in Fig.\ref{structure_nem}{\bf c}, we observed the paramagnetic phase corresponding to $h_z=0.76$, slightly beyond the transition point. Clearly, this paramagnetic phase is ``polarized''(i.e: no zero net magnetization), due to the external magnetic field.  

\begin{figure}[!htb]
 \centering
 \includegraphics[width=.35\textwidth]{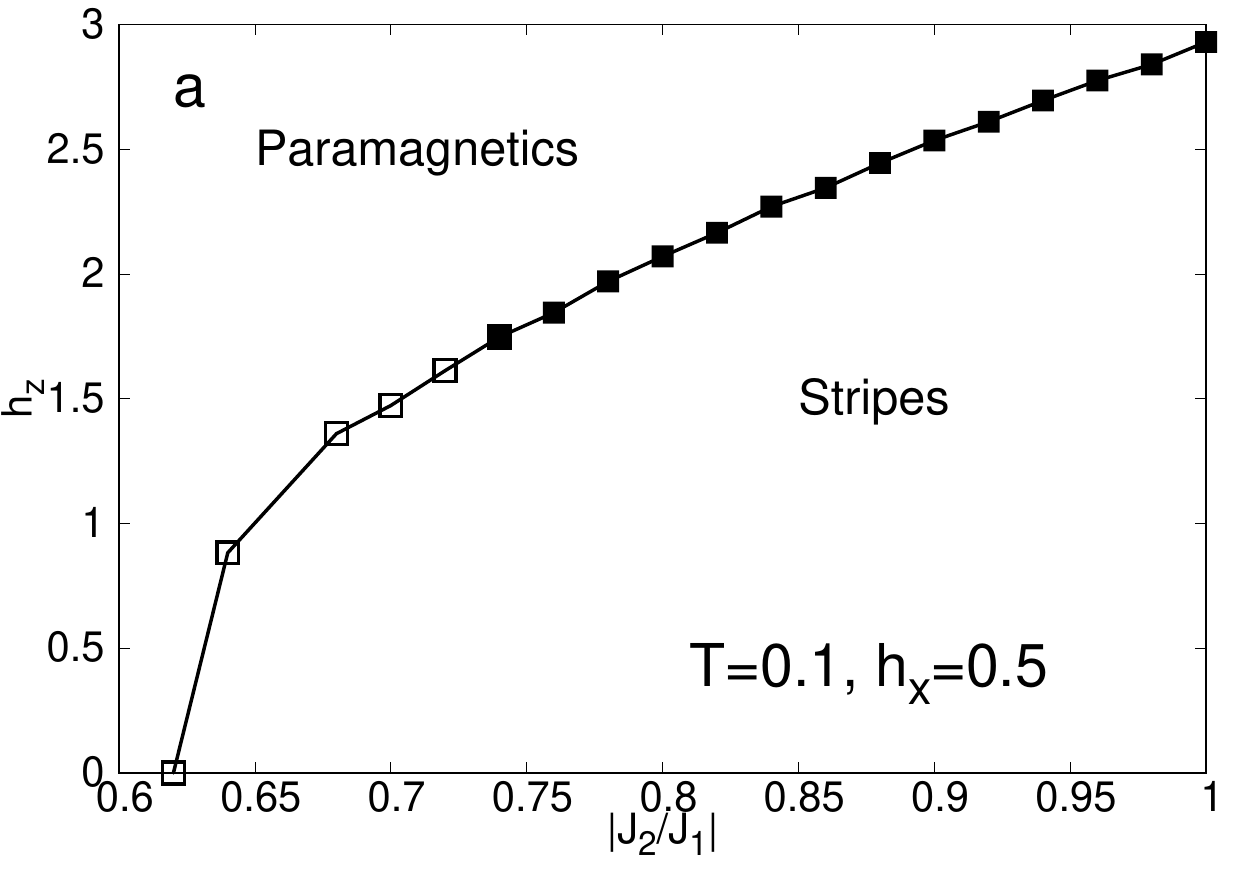}
 \includegraphics[width=.35\textwidth]{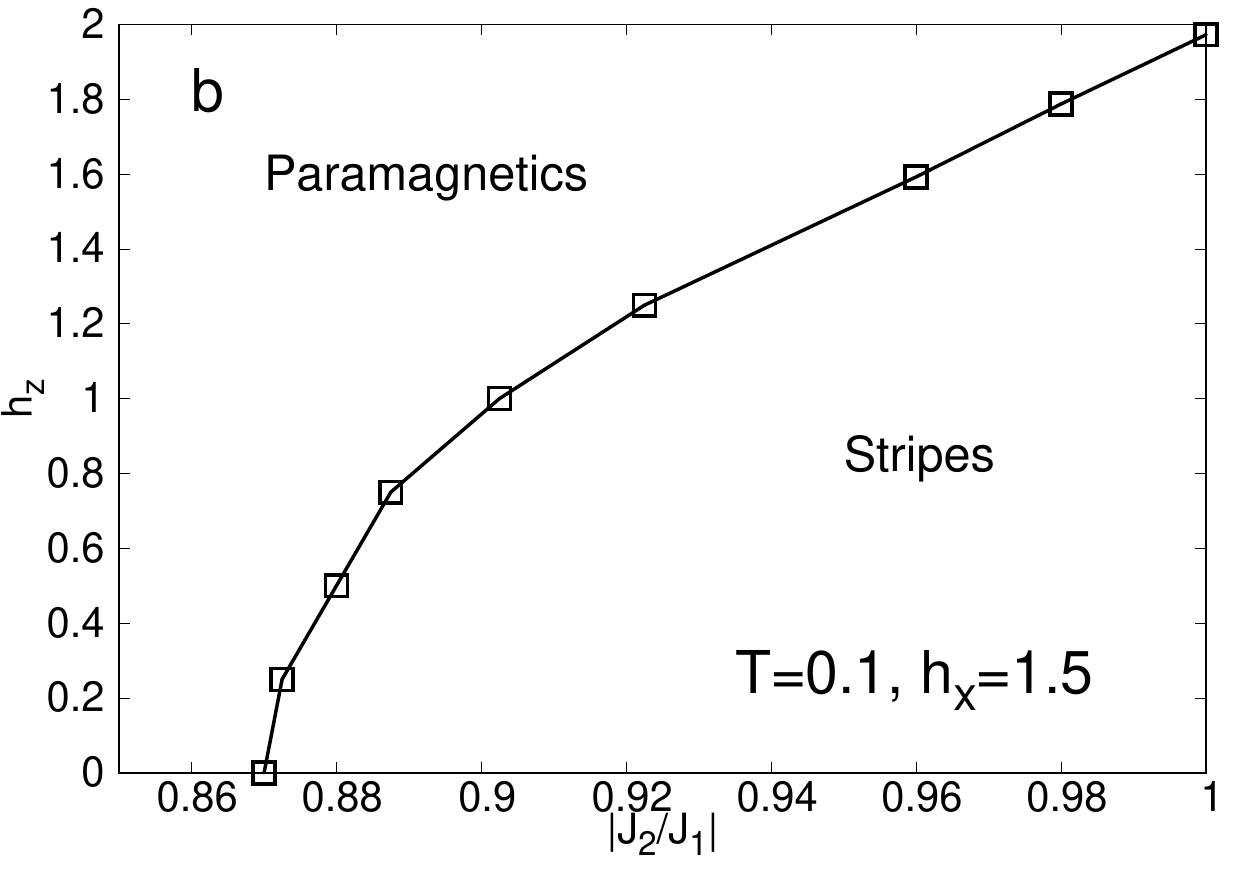}
 \includegraphics[width=.35\textwidth]{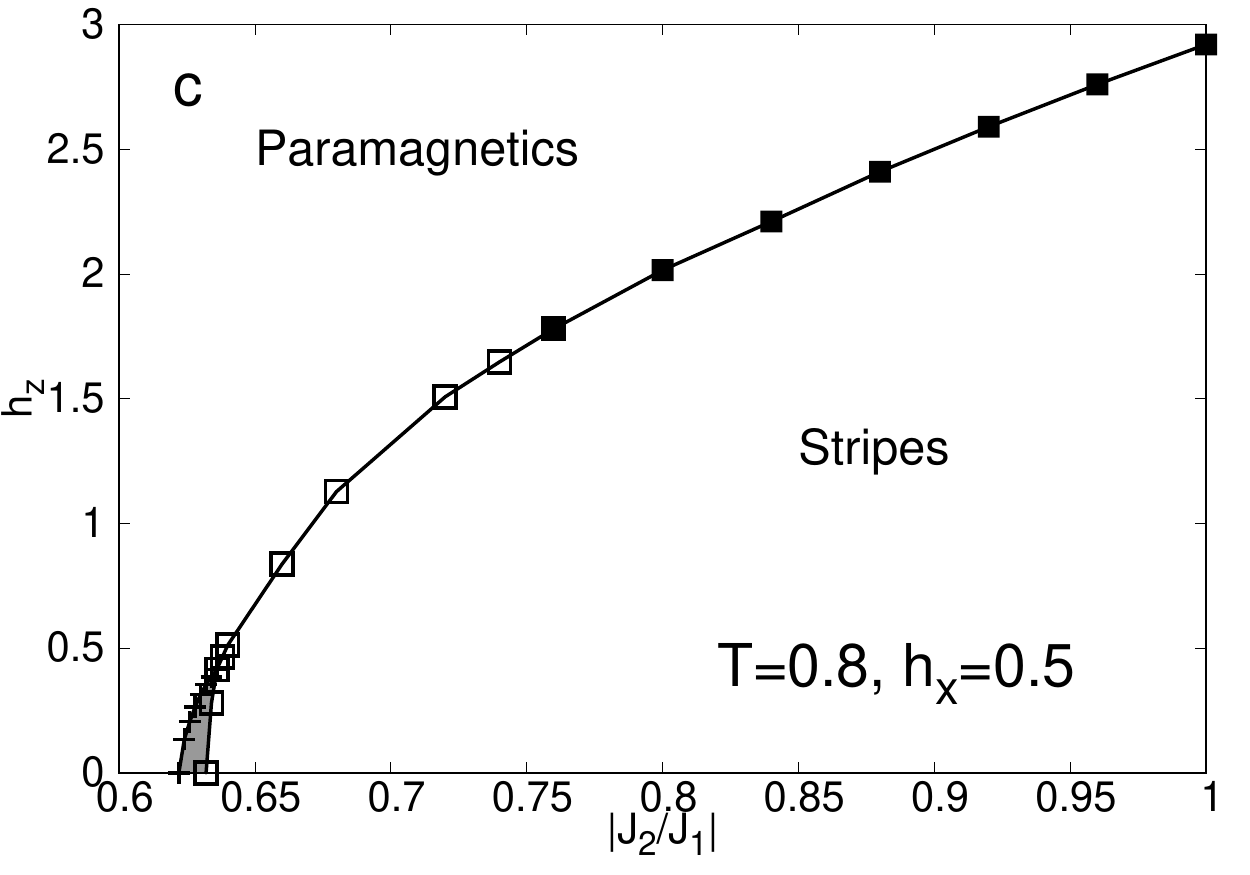}
\includegraphics[width=.35\textwidth]{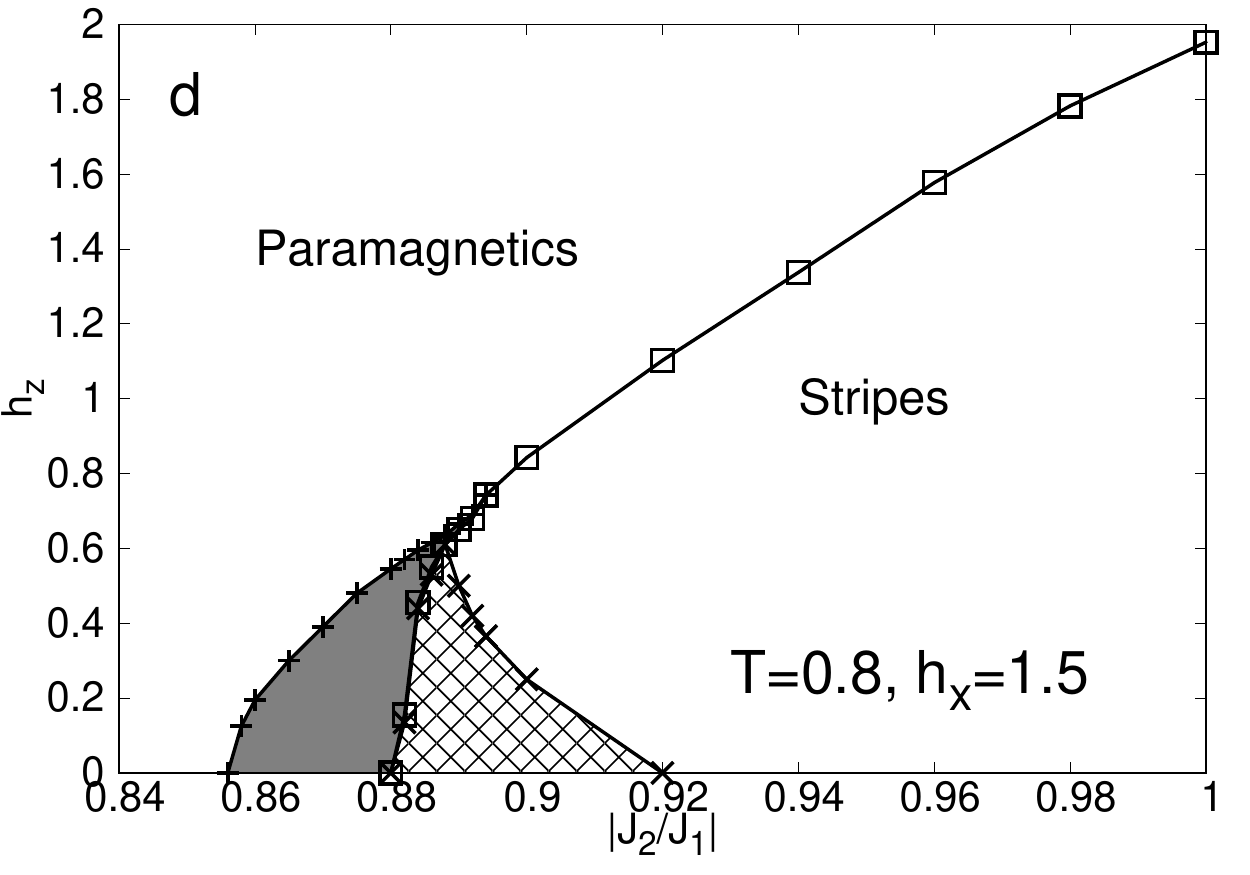}
\caption{Phase diagrams $(g,hz)$  in the presence of longitudinal field. From the top $T=0.1$, $h_x=0.5$ (panel {\bf a}), $T=0.1$, $h_x=1.5$ (panel {\bf b}), $T=0.8$, $h_x=0.5$ (panel {\bf c}), and $h_x=1.5$ (panel {\bf d}).}
\label{effect_hx_Ferro}
\end{figure}

The behaviour of  the order parameters can be condensed in the $h_z - g$ phase diagrams presented in Figure \ref{effect_hx_Ferro} for different values of $h_x$ and $T$. At low temperatures (Figure \ref{effect_hx_Ferro}{\bf a,b}) it can be observed that the longitudinal field ($h_x$) changes the character of the transition. For example, in Figure \ref{effect_hx_Ferro}({\bf a}), corresponding to a low longitudinal field, we have a line of continuous and discontinuous transitions, with a critical point separating the two behaviors while in panel {\bf b}, (larger longitudinal field) all the transition line is discontinuous. So, we are tempted to associate the application of a larger longitudinal field with an increase in the discontinuous character of the phase transition. On the other hand, at higher temperatures, $T=0.8$ (Fig.\ref{effect_hx_Ferro}{\bf c} and Fig.\ref{effect_hx_Ferro}{\bf d}), a nematic phase shows up penetrating the phase of stripes. As can be easily observed, this region broadens as we increase the value of the longitudinal field. On the other side, we can check that as frustration increase ($|J_2/J_1|$), the longitudinal field required for the occurrence of the nematic phase is larger. Notice also that in panel {\bf d}, we have a non convergence region due to numerical issues in the implementation, which is quite demanding in this region of the phase diagram.

\begin{figure}[!htb]
\centering
\includegraphics[width=.4\textwidth]{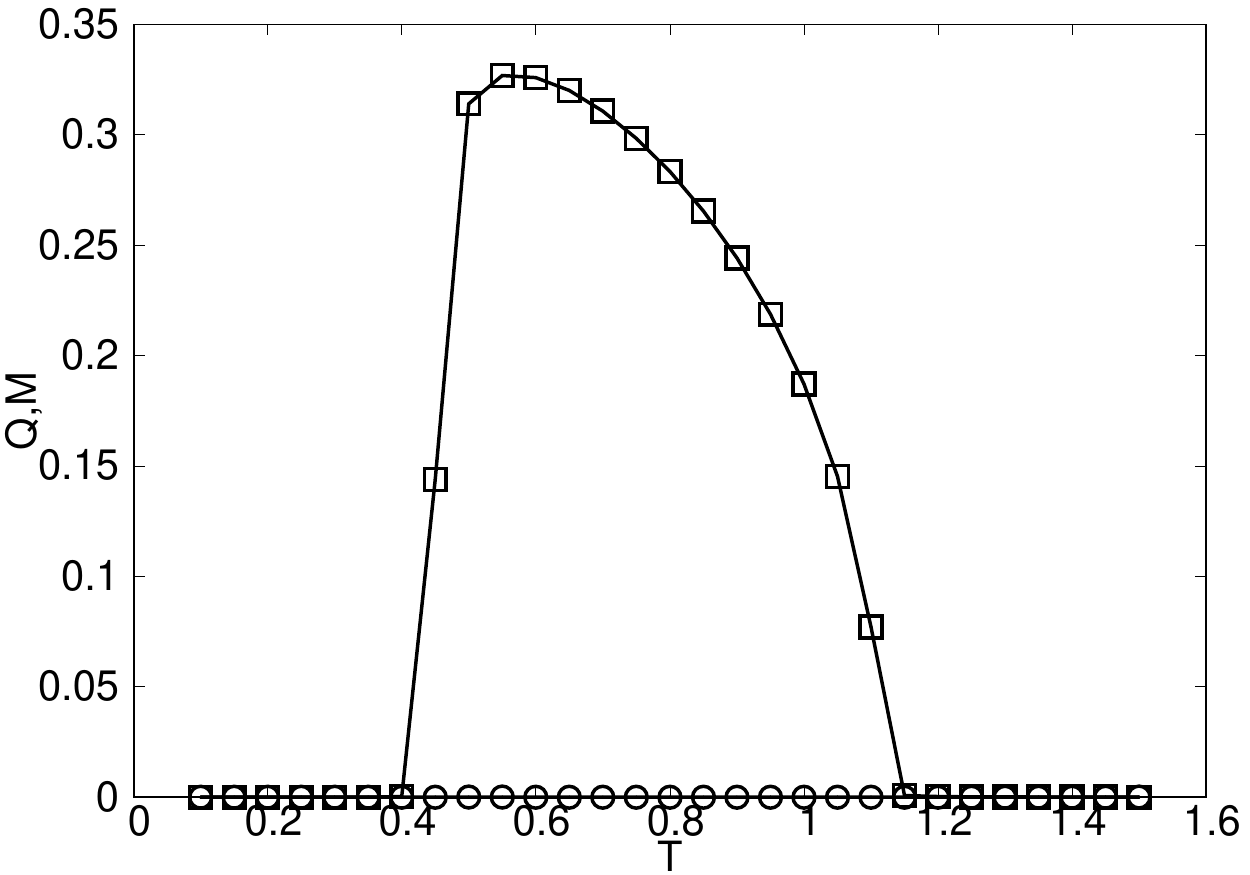}
\caption{Behavior of the orientational and translational order parameters for $h_x=2.0$, $h_z=0.1$, $J_2/J_1=-1$. A range of temperatures where nematic order develops is observed. The later certainly begins in a finite Temperature ($T\approx 0.4$).     }
\label{Order_Parameters_hx20}
\end{figure}

These results suggest that the nematic phase does not appear in a low temperature scenario in agreement with results previously observed \cite{Ours} in the classical model for $J_2/J_1=-1$. To further support this, in Fig.\ref{Order_Parameters_hx20}, we show the behavior of the order parameters for $h_x=2.0$. There we can clearly check that the nematic phase arises only  after a certain threshold in $T$.

\begin{figure}[!htb]
\centering
\includegraphics[width=.4\textwidth]{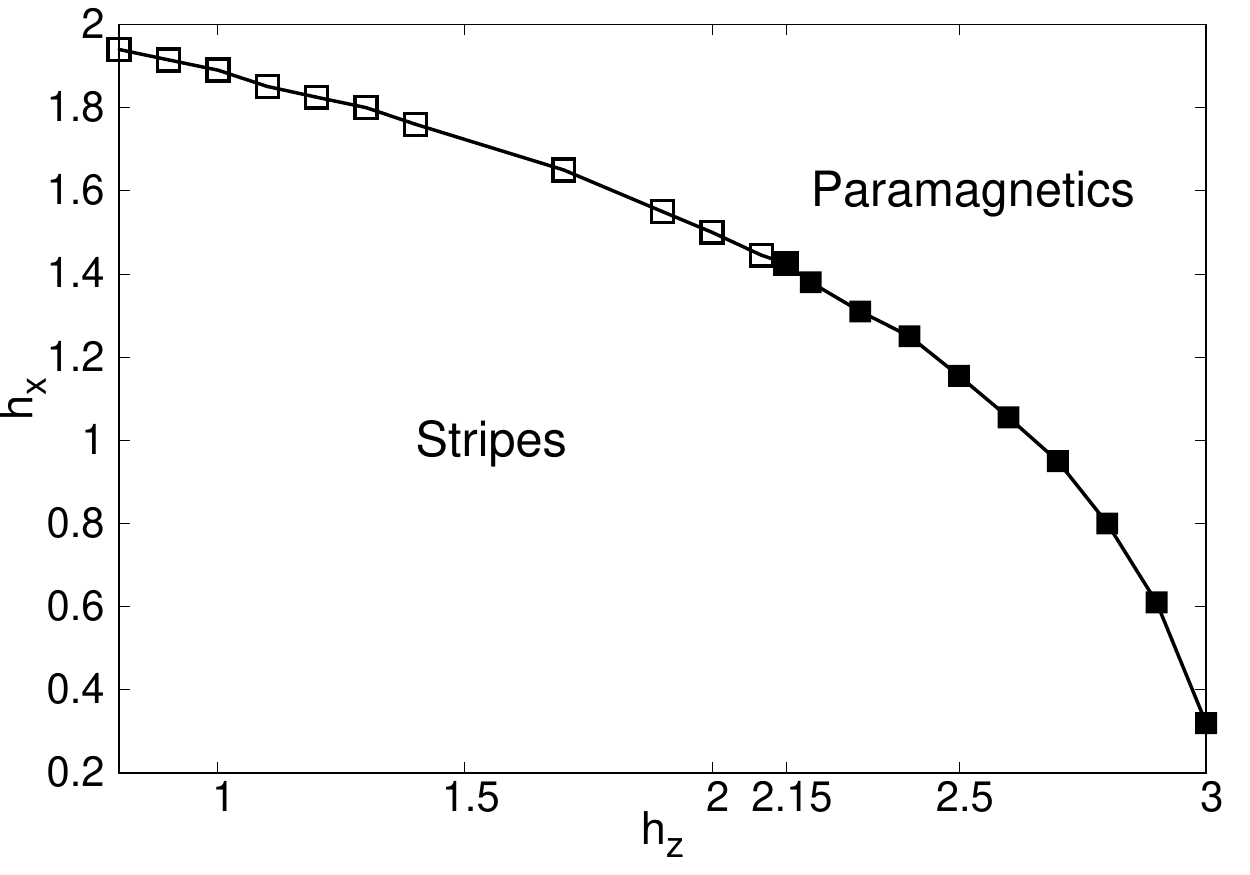}
\caption{$h_z$ $vs$ $h_x$ phase diagram at low temperatures ($T=0.1$), and $J_2/J_1=-1$. No nematic phase is present in this scenario. A critical point located around $h_z=2.15$, $h_x=1.425$ separates a region of first order phase transition of a a line of continuous transition.      }
\label{phase_diagram_hz_hx_T01}
\end{figure}

Another feature of these phase diagrams that is worth mentioning is the effect of $h_x$ in the continuity of the phase transitions. The region in which first order transition occurs widens as $h_x$ increases.  This effect goes in the opposite direction to the one we observe under the application of a transverse field. In Fig. \ref{phase_diagram_hz_hx_T01},  we show the $h_z$ $vs$ $h_x$ phase diagram for $T=0.1$, $J_2/J_1=-1$. The transition line is first order at low values of the transverse field, and large longitudinal one. In the other extreme case, a continuous transition occurs. The critical point is located around $h_z=2.15$, $h_x=1.425$.

\section{Conclusions}
In this work we studied the quantum $J_1$-$J_2$ model with nearest neighbor ferromagnetic interactions ($J_1 \ge 0$) and next nearest neighbour anti-ferromagnetic interactions ($J_2 \le 0$) in the presence of a transverse field using the Quantum Cluster Variational Method. We study the model in an extensive lattice without particular assumptions about the symmetry of the problem.  Our results show that quantum fluctuations change the order of the transition, the larger the quantum effects the wider is the range of parameters for which the transition is continuous. Moreover, quantum fluctuations may induce a gap between the ferromagnetic phase and the phase of stripes and in the presence of longitudinal fields also a pronounced nematic phase that penetrates the phase of stripes. This nematic phase is characterized by the presence of orientational order and the lack of translational order.

\subsection*{Appendix: On the convergence of QCVM near the continuous-discontinuous transition}

\begin{figure}[!htb]
\centering
\includegraphics[width=.4\textwidth]{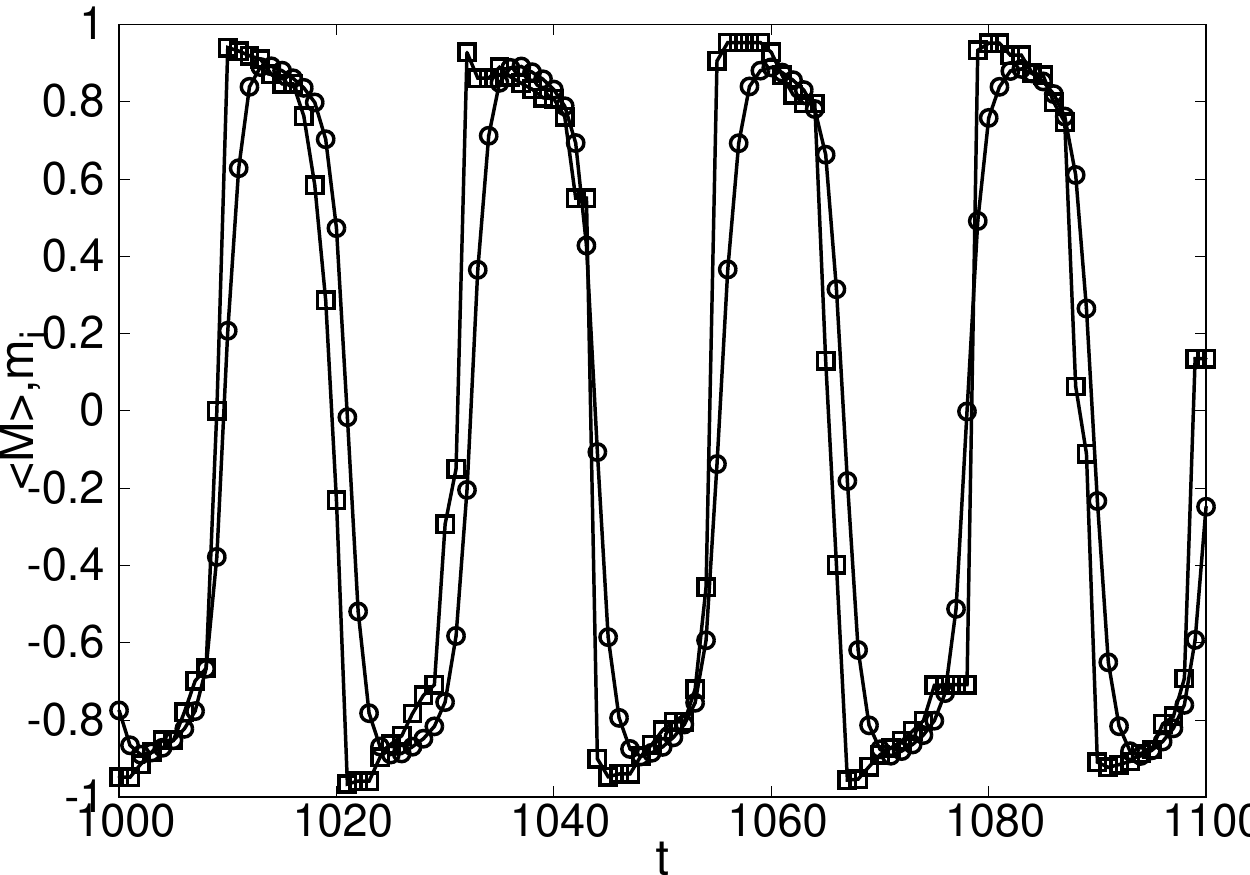}
\caption{Oscillatory behavior of the local magnetization of a single site (open squares), along with the global magnetization (open circles), in a lattice of 32 x 32 sites. The plot corresponds to the parameters $h_z=0.8$, $T=1.0$, $J_2/J_1=-0.34$, deep inside a non-convergence region. }
\label{Osc_Non_conv}
\end{figure}

 Wide non convergence regions are observed in some of the $T$-$|J_2|/J_1$ phases diagrams in Fig.\ref{effect_hz_T},  as well as in the $h_z$-$|J_2|/J_1$ in Fig.\ref{Ferro_hz}. These non-convergence regions are characterized by an oscillatory behavior of the system, which can be observed in Fig.\ref{Osc_Non_conv}. In this figure we show both the behavior of a single site's magnetization as well as the global magnetization in a 32 x 32 lattice. Both values oscillate coherently, which is the signature of a global  coordination rather than the result of single site isolated variations. An important thing to notice is that there is a decrease in the amplitude of the oscillations as we move from the ferromagnetic region to the paramagnetic one. 

\begin{figure}[!htb]
 \centering
 \includegraphics[width=.4\textwidth]{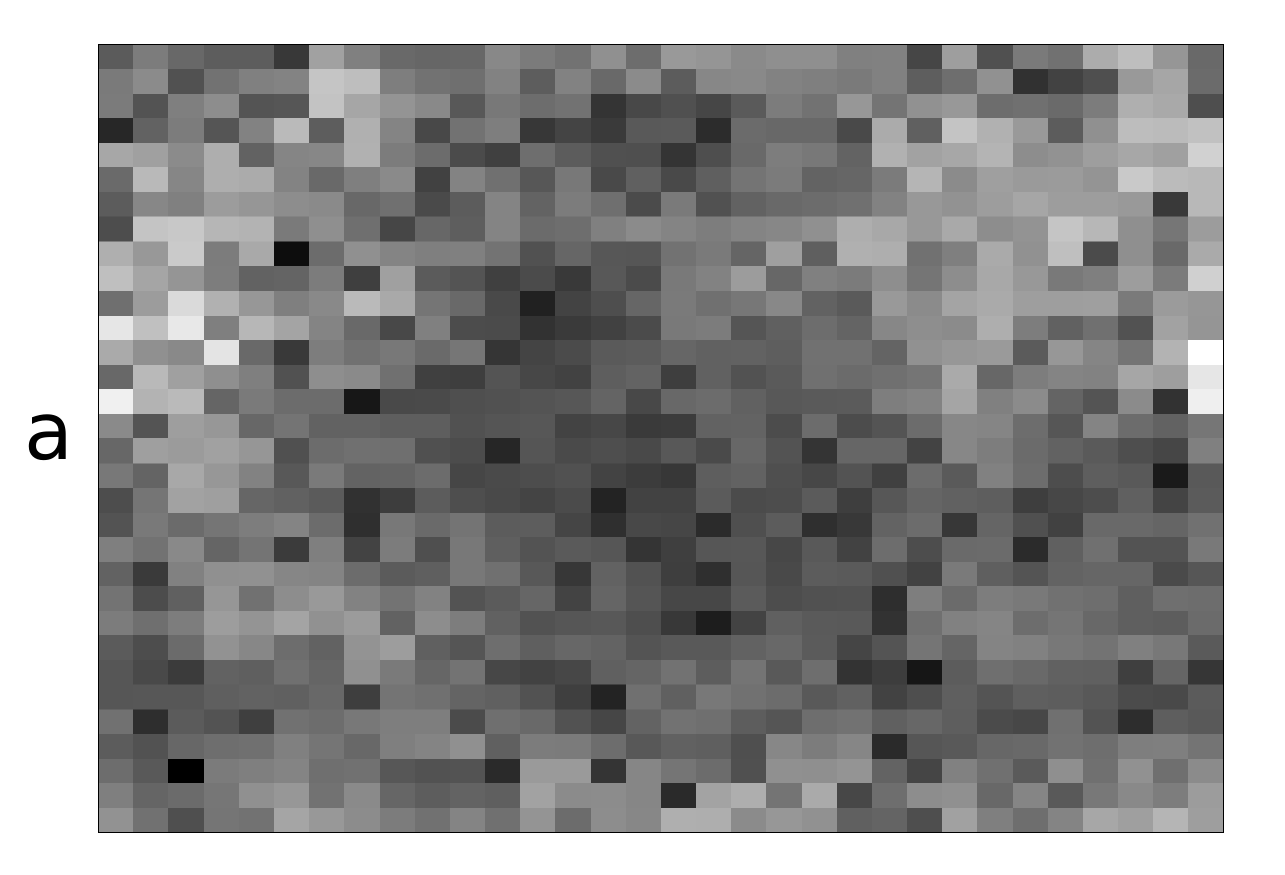}
 \includegraphics[width=.4\textwidth]{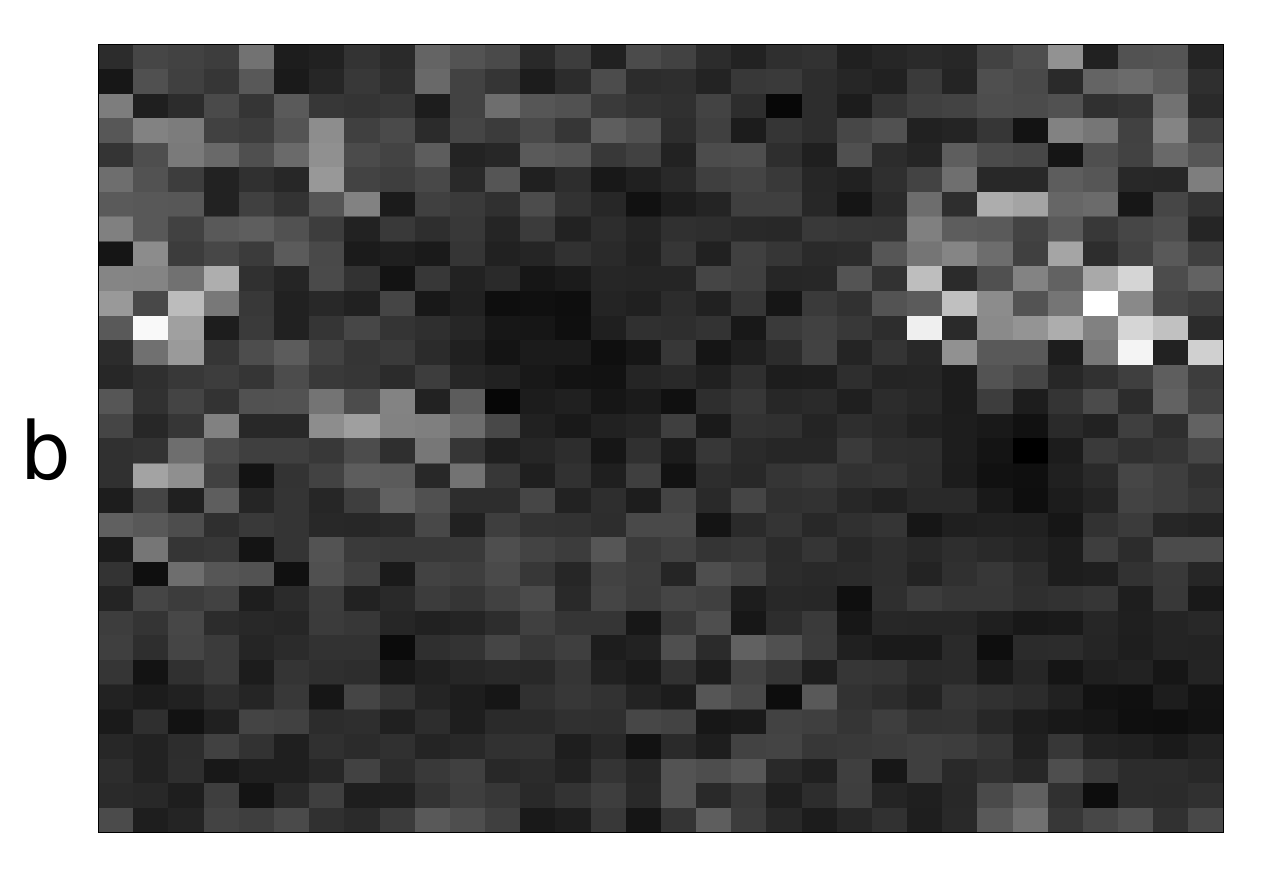}
 \includegraphics[width=.4\textwidth]{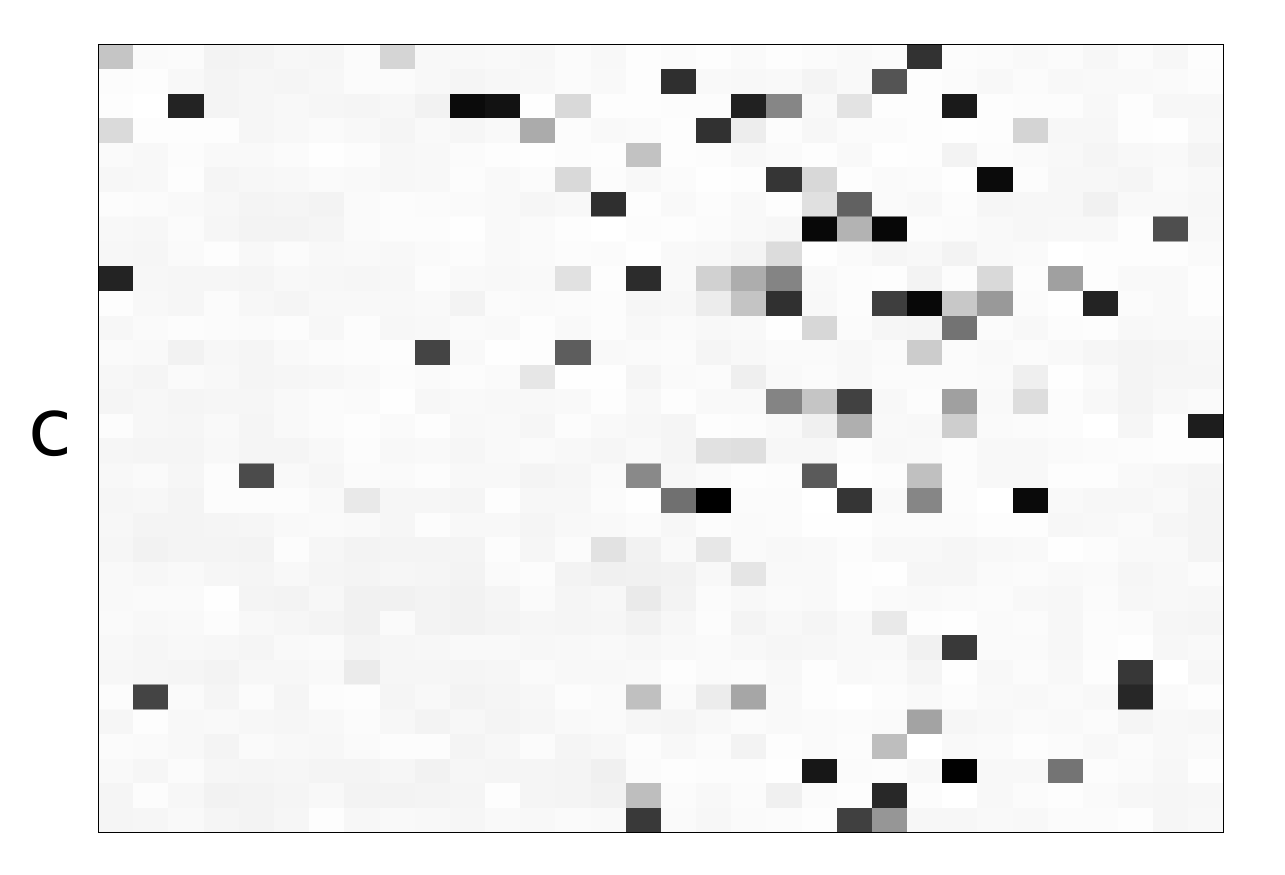}
\caption{Magnetization structure in a 32 x 32 lattice in a non convergence region for different time steps. The plot corresponds to the parameters $h_z=0.8$, $T=1.0$, $J_2/J_1=-0.34$, deep inside a non convergence region. In all the time steps a general arrange of the system favoring the mutual alignment of the spins is observed. In panel {\bf c}, the latter is particularly clear, as most of the sites show large positive magnetization. }
\label{L_32}
\end{figure}

Fig.\ref{L_32} displays the magnetization structure of a 32 x 32 lattice inside the non convergence region. According to what was previously shown in Fig.\ref{Osc_Non_conv}, there is a large coordination between most of the sites of the lattice. 

\begin{figure}[!htb]
\centering
\includegraphics[width=.4\textwidth]{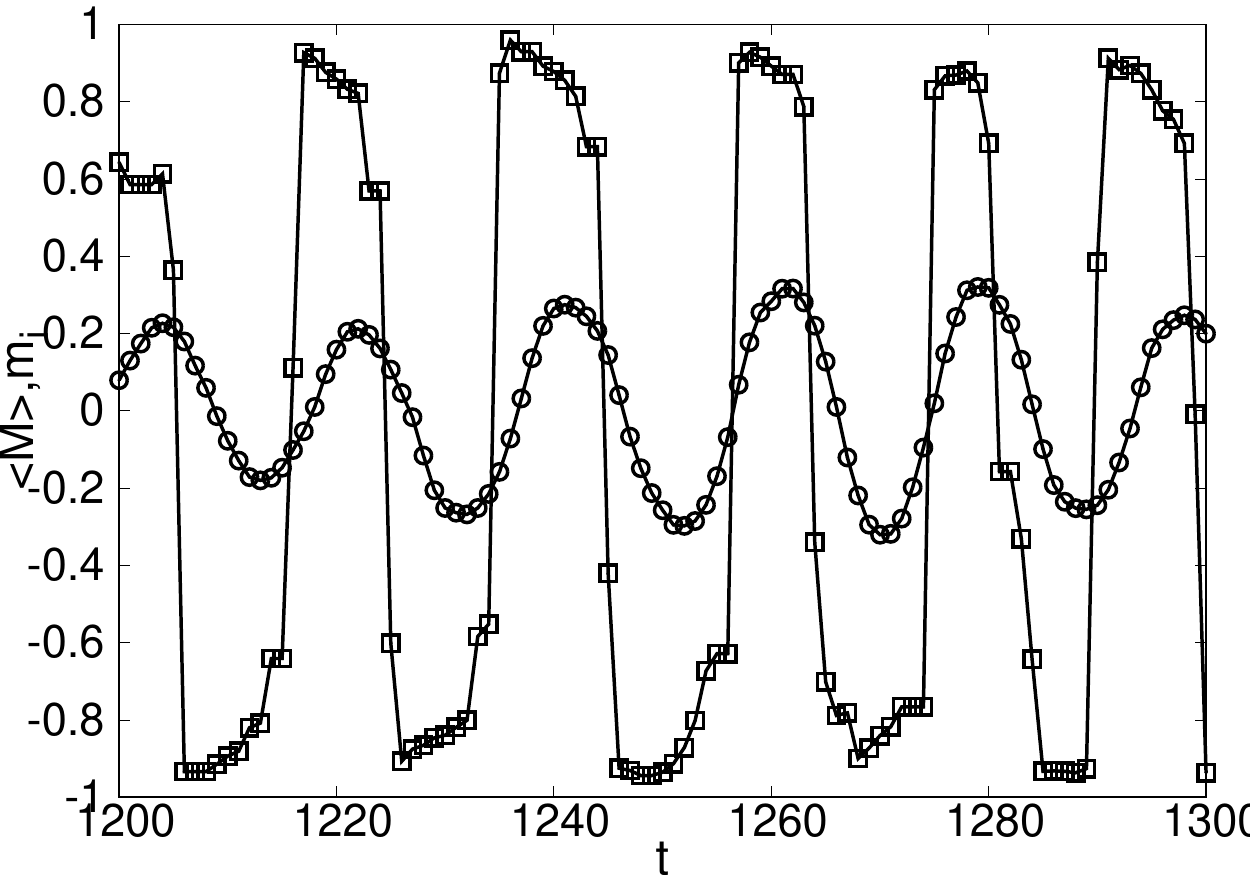}
\caption{Oscillatory behavior of the local magnetization of a single site (open squares), along with the global magnetization (open circles). The plot corresponds to the parameters $h_z=0.8$, $T=1.0$, $J_2/J_1=-0.34$, deep inside a non-convergence region in a lattice of 64 x 64 sites. }
\label{L64_osc}
\end{figure}

\begin{figure}[!htb]
\centering
\includegraphics[width=.4\textwidth]{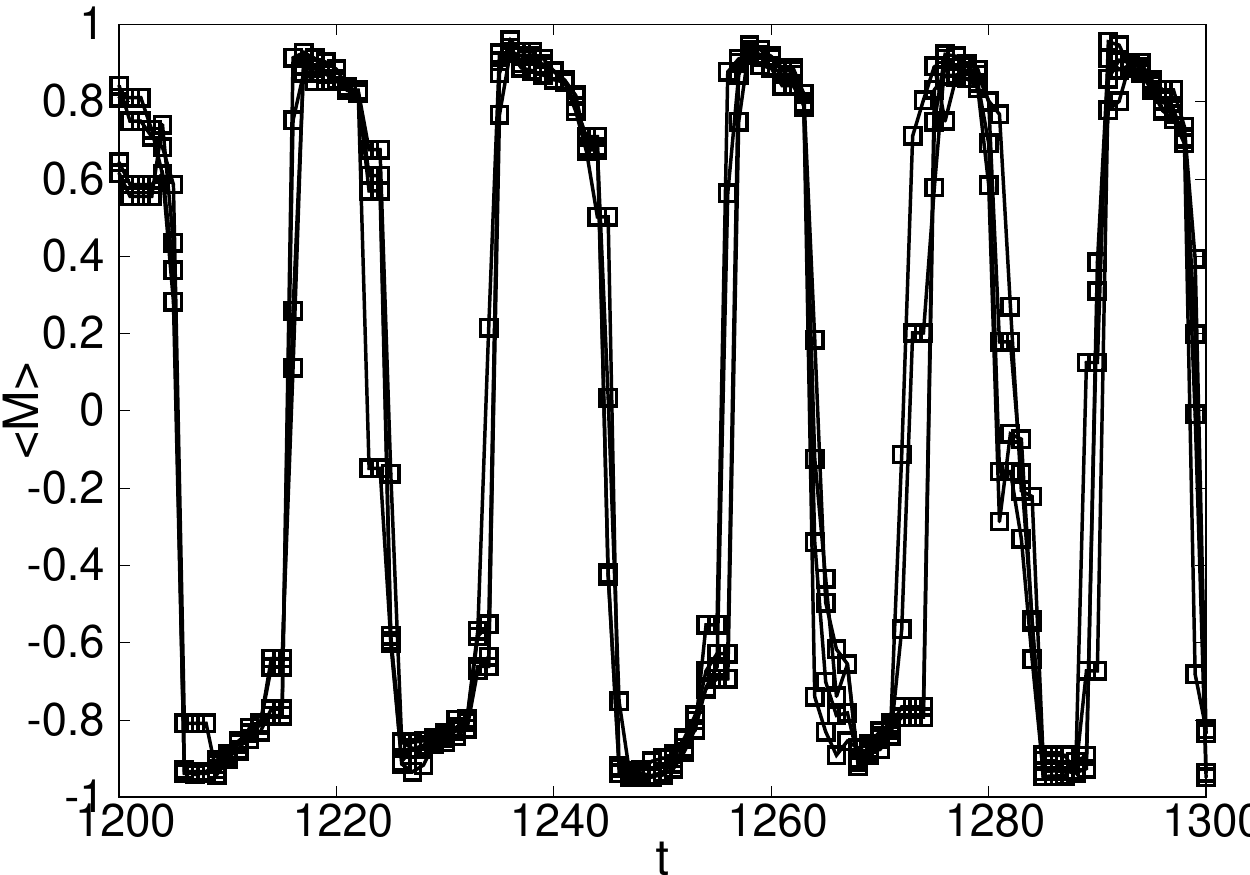}
\caption{Oscillatory behavior of the local magnetization of four sites close to one another, in a lattice of 64 x 64 sites. The plot corresponds to the parameters $h_z=0.8$, $T=1.0$, $J_2/J_1=-0.34$, deep inside a non-convergence region. }
\label{near_sites_osc}
\end{figure}

\begin{figure}[!htb]
 \centering
 \includegraphics[width=.4\textwidth]{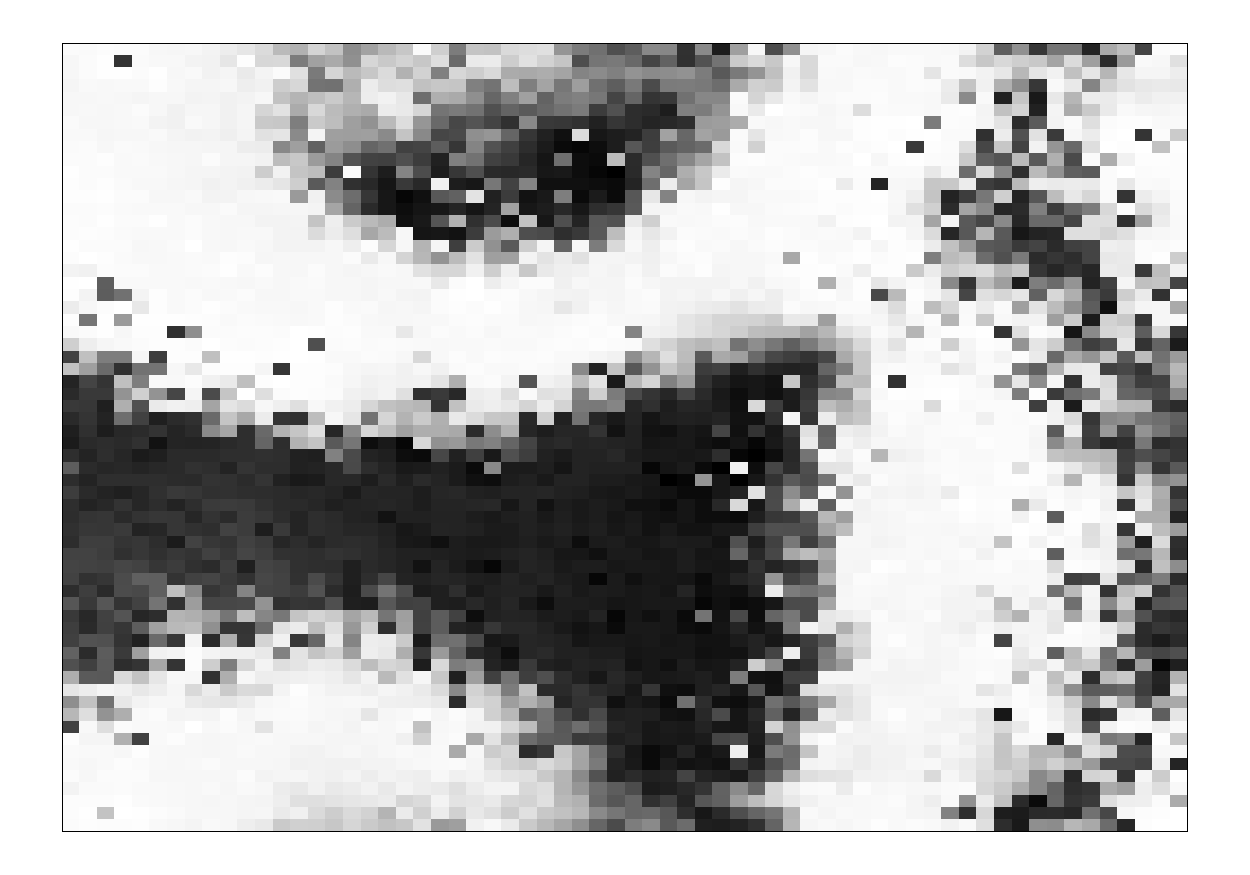}
 \includegraphics[width=.4\textwidth]{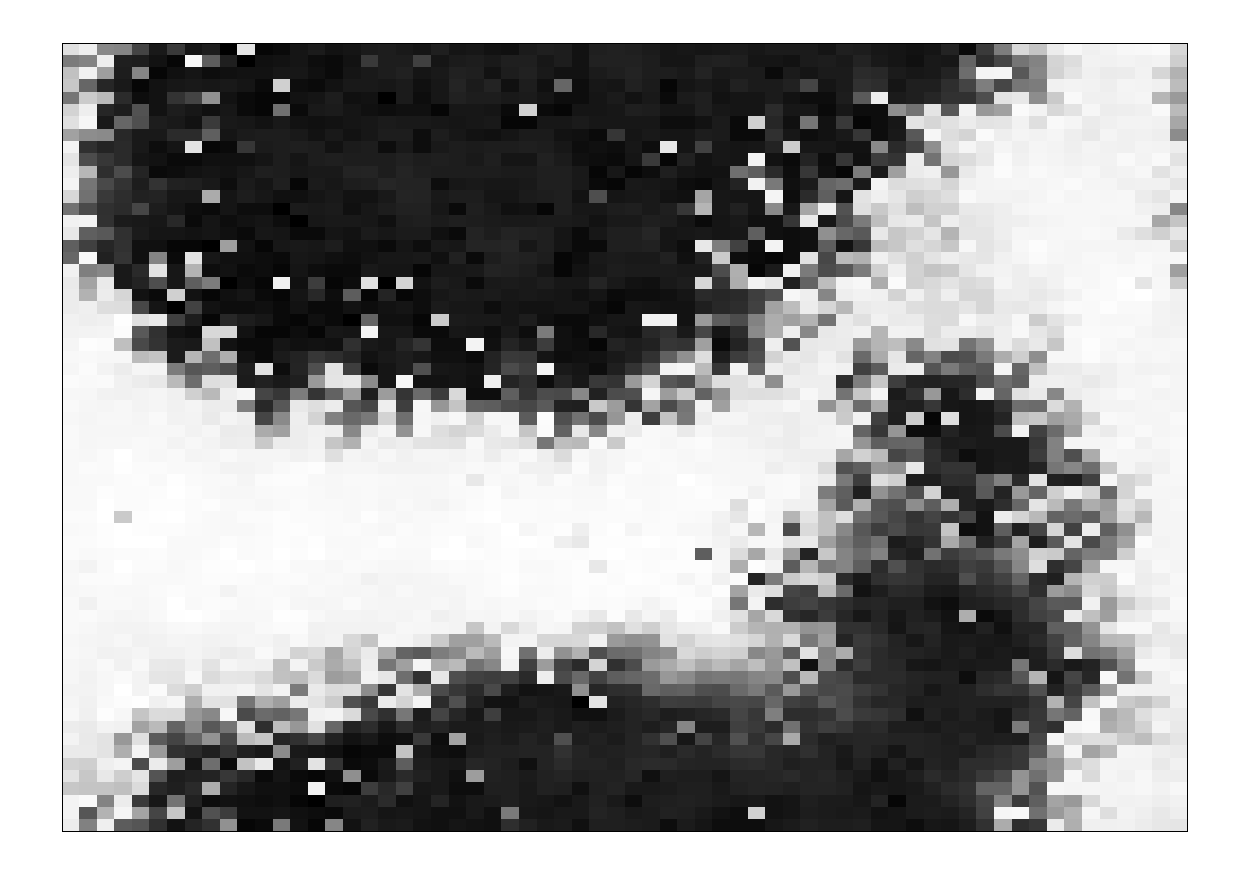}
\caption{Cluster formation in a 64 x 64 lattice in a non convergence region for two different time steps. The plot corresponds to the parameters $h_z=0.8$, $T=1.0$, $J_2/J_1=-0.34$, $h_x=0$ deep inside a non convergence region. }
\label{clusters}
\end{figure}

For much larger regions, say $L=64$, both the global and local magnetization keep oscillating coherently, yet the global magnetization does it with a shorter amplitude (Fig.\ref{L64_osc}). 
This effect can be linked to the formation of clusters. In order to further motivate this statement we first show in Fig.\ref{near_sites_osc} the oscillations of the magnetization of a site chosen randomly and some of its neighbors. We can observe what can be described as a local coordination. 
Finally, in Fig.\ref{clusters}, we show the 64 x 64 lattice structure for two different time steps. It can be directly observed in the two cases, the formation of clusters of sites pointing in the same direction, which visually show what we supposed was the reason for the apparent contradiction in the oscillation of the local and global magnetization.

\bibliographystyle{apsrev4-1}%{plain}
\bibliography{bibqcvm}

\end{document}